\documentclass[twocolumn]{aastex62}
\usepackage{graphicx,url}
\usepackage{amsmath,amssymb}
\usepackage{hyperref}
\hypersetup{colorlinks,
	linkcolor=blue, 
	urlcolor=magenta, 
	anchorcolor=blue,
	citecolor=blue
}
\usepackage{bm}
\usepackage{xcolor}
\usepackage{mathrsfs}
\usepackage{natbib}

\begin{document}

\title{\large\bfseries ECoPANN: A Framework for Estimating Cosmological Parameters \\ using Artificial Neural Networks}

\correspondingauthor{Jun-Qing Xia}
\email{xiajq@bnu.edu.cn}

\author{Guo-Jian Wang}
\affil{Department of Astronomy, Beijing Normal University, Beijing 100875, China}

\author{Si-Yao Li}
\affil{SenseTime Research, Beijing 100080, China}

\author{Jun-Qing Xia}
\affil{Department of Astronomy, Beijing Normal University, Beijing 100875, China}

\begin{abstract}
In this work, we present a new method to estimate cosmological parameters accurately based on the artificial neural network (ANN), and a code called ECoPANN (Estimating Cosmological Parameters with ANN) is developed to achieve parameter inference. We test the ANN method by estimating the basic parameters of the concordance cosmological model using the simulated temperature power spectrum of the cosmic microwave background (CMB). The results show that the ANN performs excellently on best-fit values and errors of parameters, as well as correlations between parameters when compared with that of the Markov Chain Monte Carlo (MCMC) method. Besides, for a well-trained ANN model, it is capable of estimating parameters for multiple experiments that have different precisions, which can greatly reduce the consumption of time and computing resources for parameter inference. Furthermore, we extend the ANN to a multibranch network to achieve a joint constraint on parameters. We test the multibranch network using the simulated temperature and polarization power spectra of the CMB, Type Ia supernovae, and baryon acoustic oscillations, and almost obtain the same results as the MCMC method. Therefore, we propose that the ANN can provide an alternative way to accurately and quickly estimate cosmological parameters, and ECoPANN can be applied to the research of cosmology and even other broader scientific fields.
\end{abstract}
\keywords{Cosmological parameters (339); Observational cosmology (1146); Computational methods (1965); Astronomy data analysis (1858); Neural networks (1933)}

\section{\bf Introduction}\label{sec:introduction}

The improvement of the quality and sensitivity of the cosmic microwave background (CMB; \citet{WMAP,Planck2018:VI}) observation ushered the research of cosmology into the precision era. The CMB was accurately observed by many space-based, ground-based, and suborbital experiments, and it is a very powerful way for us to study the universe. The statistical properties of the CMB are coincident with the predictions of the six-parameter standard $\Lambda$CDM cosmological model \citep{Planck2018:VI}. In the $\Lambda$CDM model, parameters are tightly constrained owing to the high precision of the observation of the CMB. In terms of constraining cosmological parameters, the Markov Chain Monte Carlo (MCMC) technique is widely used by scientists in this field for its excellent performance. 

However, when confronting more parameters and large amounts of data, MCMC will consume quantities of time and computing resources. Therefore, new methods and techniques are needed to analyze a huge amount of data in the present and future astronomy. In order to solve this problem, \citet{Auld:2007,Auld:2008} presented a Bayesian inference algorithm called {\sc CosmoNet}, which is based on training an artificial neural network (ANN), to accelerate the calculation of CMB power spectra, matter power spectra, and likelihood functions for use in cosmological parameter estimation. Furthermore, \citet{Graff:2012} presented the blind accelerated multimodal Bayesian inference (BAMBI), an algorithm for rapid Bayesian analysis that combines the benefits of nested sampling and ANNs, to learn the likelihood function.

The ANN, composed of linear and nonlinear transformations of input variables, has been proven to be a ``universal approximator'' \citep{Cybenko:1989,Hornik:1991}, which can represent a great variety of functions. This powerful property of the ANN allows its wide use in regression and estimation tasks. With the development of computer hardware in the past decade, the ANN is now capable of containing deep layers and training with a large amount of data. Recently, methods based on ANNs have outstanding performances in solving cosmological problems in both accuracy and efficiency. For example, it performs excellently in analyzing gravitational wave \citep{George:2018a,George:2018b,George:2018c,Shen:2019,LiXiangru:2020}, estimating parameters of 21 cm signal \citep{Shimabukuro:2017,Schmit:2018}, discriminating the cosmological and reionization models \citep{Schmelzle:2017,Hassan:2018}, searching and estimating parameters of strong gravitational lenses \citep{Jacobs:2017,Petrillo:2017,Hezaveh:2017,Pourrahmani:2018,Schaefer:2018}, classifying the large-scale structure of the universe \citep{Aragon-Calvo:2019}, estimating cosmological parameters \citep{Fluri:2018,Fluri:2019,Ribli:2019,Ntampaka:2020}, studying the evolution of dark energy models \citep{Escamilla-Rivera:2020}, and reconstructing functions from cosmological observational data \citep{Wanggj:2020a,Wanggj:2020b}.

In this work, we show that the ANN is capable of estimating cosmological parameters with high accuracy, which makes the ANN an alternative to the MCMC method in parameter estimation. We test the ANN method by constraining parameters of the $\Lambda$CDM model with the simulated data sets of the CMB, Type Ia supernovae (SNe Ia), and baryon acoustic oscillations (BAOs). Based on PyTorch\footnote{\url{https://pytorch.org/docs/master/index.html}}, an open-source optimized tensor library for deep learning, we have developed a code, called Estimating Cosmological Parameters with ANN (ECoPANN\footnote{\url{https://github.com/Guo-Jian-Wang/ecopann}}), to estimate parameters in our analysis. It should be noted that the algorithm ECoPANN is different from the previous cosmological Bayesian inference algorithms {\sc CosmoNet} and BAMBI. Both {\sc CosmoNet} and BAMBI adopted ANNs to replace parts of the calculation of MCMC procedure. Thus, both of them are still working based on the MCMC method. However, ECoPANN is designed to estimate parameters directly from the observational data sets, which is a fully ANN-based framework that is different from the Bayesian inference.

This paper is organized as follows: In section \ref{sec:method}, we illustrate the method of estimating parameters, which contains the introduction to the ANN, hyperparameters of the ANN, and training and parameter inference using the ANN. Section \ref{sec:application_CMB} shows the application of the ANN method to the CMB experiments. Section \ref{sec:joint_constraint} presents a joint constraint on parameters with multibranch network. Section \ref{sec:test_hyperparameters} shows the effect of hyperparameters of the ANN on the parameter estimation. In section \ref{sec:discussion}, discussions about the ANN method in parameter estimation are presented. Finally, conclusions are shown in section \ref{sec:conclusions}.

\section{\bf Method}\label{sec:method}

In this section, we will first introduce the ANN method, then the settings of hyperparameters of the ANN, and finally the process of training the ANN and parameter inference.

\subsection{Artificial Neural Networks}\label{sec:ANN}

\begin{figure}
	\centering
	\includegraphics[width=0.45\textwidth, angle=0]{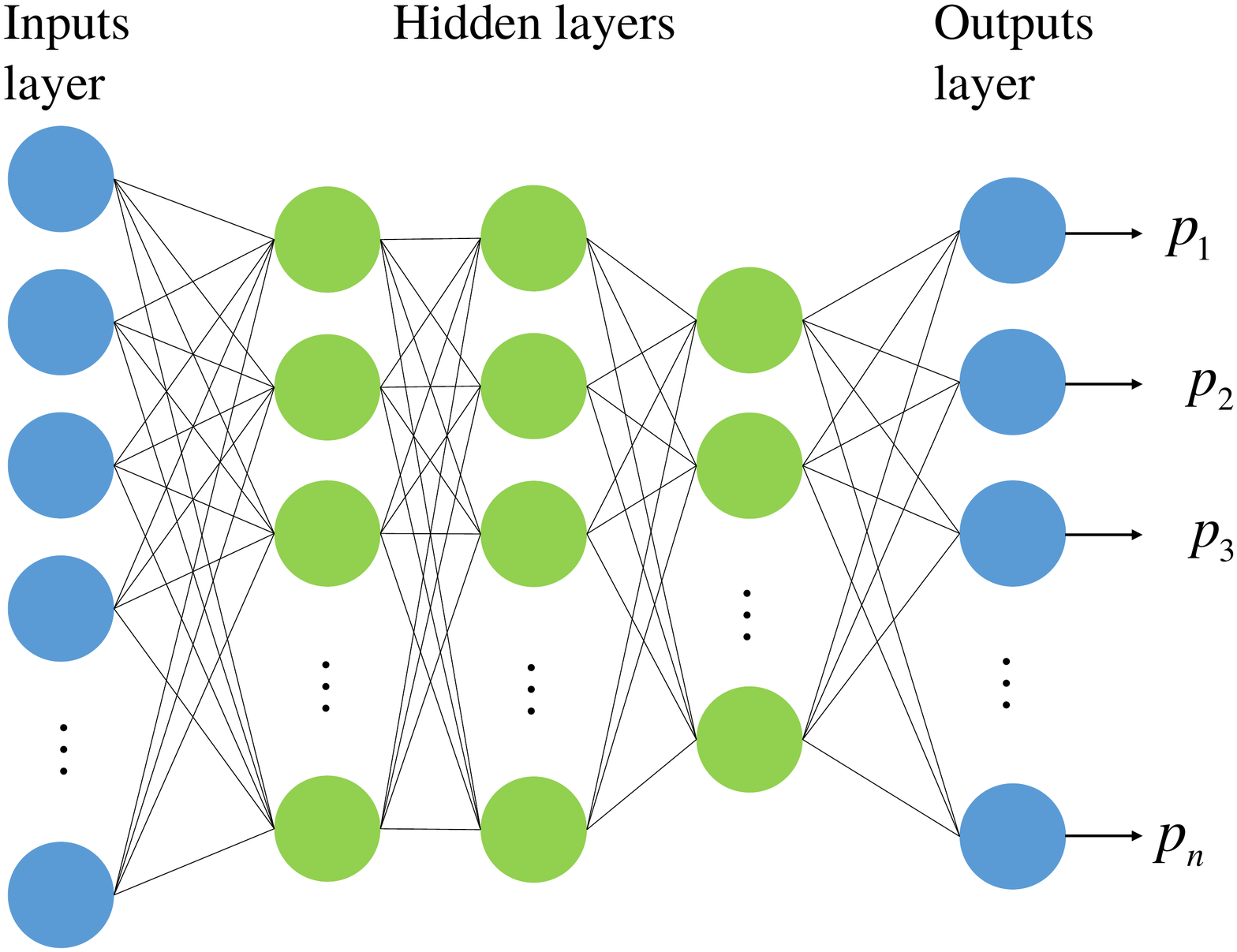}
	\caption{General structure of an ANN. The input is the observational data, and the outputs are parameters of a specific cosmological model.}\label{fig:nn_model}
\end{figure}

An ANN, also called a neural network (NN), is a mathematical model that is inspired by the structure and functions of biological NNs, and it generally consists of an input layer, hidden layers, and an output layer. In Figure \ref{fig:nn_model}, we show a general structure of the ANN. For the task of estimating cosmological parameters, the observational data are fed to the input layer, then the information of observational data passes through each hidden layer, and finally the cosmological parameters are output from the output layer. Specifically, each layer accepts a vector, the elements of which are called neurons, from the former layer as input, then applies a linear transformation and a nonlinear activation on the input, and finally propagates the current result to the next layer. Formally, in a vectorized style,
\begin{equation}\label{eq:def1}
\bm z_{i+1} = \bm x_{i}W_{i+1} + \bm b_{i+1},
\end{equation}
\begin{equation}\label{eq:def2}
\bm x_{i+1} = f(\bm z_{i+1}),
\end{equation}
where $\bm{x}_{i}$ is the input row vector of the $i$th layer, $W_{i+1}$ and $\bm b_{i+1}$ are linear weights and biases to be learned, $\bm z_{i+1}$ is the intermediate vector after linear transformation, and $f$ is the element-wise nonlinear function (also known as activation function). The output layer only takes linear transformation. Here we take the randomized leaky rectified linear units (RReLU; \citet{RReLU}) as the activation function, which has the form
\begin{equation}
f(x) = \left\{\begin{matrix}
x & \text{if } x \geq 0 \\
ax & \text{if } x < 0,
\end{matrix}\right.
\end{equation}
where $a$ is a random number sampled from a uniform distribution $U(l, u)$, and $l, u \in [0, 1)$. Here we adopt the default settings of $l=1/8$ and $u=1/3$ in Pytorch.

ANNs are usually designed to process a batch of data simultaneously. Therefore, as a hyperparameter, batch size is usually used in the ANN, which defines the number of samples that propagate through the network in one iteration. Consider a matrix $X\in \mathbb{R}^{m\times n}$, where $m$ is the batch size and each row of $X$ is an independent input vector, and $n$ is the length of the input vector (for the input layer, $n$ equals the number of observational data points); then, Equations (\ref{eq:def1}) and (\ref{eq:def2}) are replaced by the following batch-processed version:
\begin{equation}\label{eq:matrixdef1}
Z_{i + 1} = X_{i}W_{i+1}+B_{i + 1},
\end{equation}
\begin{equation}\label{eq:matrixdef2}
X_{i+1} = f(Z_{i+1}),
\end{equation}
where $B_{i+1}$ is the vertically replicated matrix of $\bm b_{i+1}$ in Equation \eqref{eq:def1}. An ANN equals a function $f_{W, b}$ on input $X$. In supervised learning tasks, every input datum is labeled corresponding to a ground-truth target $Y \in \mathbb{R}^{m\times p}$, where $p$ is the length of the output vector (also equal to the number of the cosmological parameters). The purpose of training an ANN is to minimize the difference between the predicted result $\hat{Y} = f_{W, b}(X)$ and the ground truth, which is quantitatively mapped with a loss function $\mathcal{L}$, by optimizing the parameters $W$ and $\bm b$. We take the least absolute deviation as the loss function, which has the following form:
\begin{equation}\label{eq:L1_loss}
\mathcal L = \frac{1}{mp} ||\hat{Y} - Y||,
\end{equation}
where the losses divided by $m$ and $p$ mean that they are averaged over cosmological parameters and also averaged over samples in the minibatch.

Following the differential chain rule, one could backward manipulate gradients of parameters in the $i$th layer from the $(i+1)$th layer, which is well recognized as the back-propagation algorithm. Formally, in a vectorized batch style \citep{LeCun:2012},
\begin{align}
\frac{\partial \mathcal L}{\partial Z_{i + 1}}&=f^{\prime}(Z_{i + 1})\frac{\partial \mathcal L}{\partial X_{i+1}},\\
\frac{\partial \mathcal L}{\partial W_{i + 1}}&=X_{i}^T\frac{\partial \mathcal L}{\partial Z_{i+1}},\\
\frac{\partial \mathcal L}{\partial X_{i}}&= W_{i+1}^T \frac{\partial \mathcal L}{\partial Z_{i + 1}}, \\
\frac{\partial \mathcal L}{\partial B_{i+1}}&=\frac{\partial \mathcal L}{\partial Z_{i+1}}.
\end{align}
where operator $\frac{\partial \mathcal L}{\partial \cdot}$ represents element-wise partial derivatives of $\mathcal L$ on corresponding indices, and $f^\prime$ is the derivative of the nonlinear function $f$. The network parameters are then updated by a gradient-based optimizer in each iteration. Here, we adopt Adam \citep{Kingma:2014} as the optimizer, which can accelerate the convergence.

In addition, the batch normalization, which is proposed by \cite{Ioffe:2015}, is implemented before every nonlinear layer. Batch normalization is tested to stabilize the distribution among variables; hence, it benefits the optimization and accelerates the convergence, and it also enables us to use higher learning rates and care less about initialization.

\subsection{Hyperparameters}\label{sec:hyperparameters}

There are many hyperparameters that should be selected before using ANNs for parameter estimations, such as the number of hidden layers, the number of neurons in each layer, learning rate, batch size, activation function, and loss function. Some of them are fixed in ECoPANN, and some are optimal. Here we illustrate the setting of hyperparameters in ECoPANN, and we will test the effect of some hyperparameters on the parameter estimation in section \ref{sec:test_hyperparameters}.

There is no suitable theory for determining the most appropriate network structure for a specific task. In general, the structure of an ANN is determined by experience. In our analysis, we take an ANN model with three hidden layers, as shown in Figure \ref{fig:nn_model}. Moreover, we design a model architecture that the number of neurons in each hidden layer is decreased proportionally. Specifically, the number of neurons in the $i$th hidden layer is
\begin{equation}\label{equ:neuron_num}
N_i=\frac{N_{\rm in}}{F^{i}},
\end{equation}
where $N_{\rm in}$ is the number of neurons of the input layer and $F$ is the decreasing factor of the number of neurons, which is defined by
\begin{equation}
F=\left(\frac{N_{\rm in}}{N_{\rm out}}\right)^{\frac{1}{n+1}},
\end{equation}
where $N_{\rm out}$ is the number of neurons of the output layer and $n$ is the number of hidden layers. Due to the decreasing factor, the number of neurons in the $i$th hidden layer may not be an integer; thus, in the actual calculations, $N_i$ should be rounded to an integer. Note that $N_{\rm in}$ and $N_{\rm out}$ are determined by the number of observational data points and those of the cosmological parameters to be estimated. Thus, the number of neurons in each layer is totally determined by the observational data and cosmological parameters.

Learning rate is a hyperparameter that controls how much to adjust the weights and biases (Equation \ref{eq:def1}) of the ANN with respect to the loss gradient, usually in the range between 0 and 1. Here, the learning rate is initially set to $10^{-2}$ and decreases with the number of epochs to $10^{-8}$. The batch size is set according to the number of the training samples to ensure that there are four iterations at each epoch. We set the number of epochs to $2\times10^3$; thus, the total number of iterations is $8\times10^3$, which is large enough to ensure that the loss function no longer decreases.

\subsection{Training and Parameter Inference}\label{sec:training_parameterInference}

In the process of estimating cosmological parameters with ECoPANN, the ANN is firstly trained with data simulated by the cosmological model, and then cosmological parameters can be determined by the trained ANN model. In this section, we illustrate the process of training the ANN and the method of estimating cosmological parameters with the ANN.

\subsubsection{Training Set}\label{sec:get_training_set}

The ANN aims to make a mapping from the input data to the output data; thus, for the task of parameter inference, the ANN actually learns a mapping between the measurement and the corresponding cosmological parameters. Therefore, in order to enable the trained ANN to have a reasonable prediction for the observational data, the parameter space of the training set should be large enough to cover the true values of parameters of the observational data. In our analysis, we set the range of each parameter to $[P-5\sigma_p, P+5\sigma_p]$, where $P$ is the mean of the posterior distribution of the parameter and $\sigma_p$ is the corresponding $1\sigma$ error. This parameter space is large enough to cover the posterior distribution of the parameters. In the parameter space, the cosmological parameters of the training set are simulated according to the uniform distribution.

\subsubsection{Add Noise}\label{sec:add_noise}

In supervised learning tasks, training sets are generally expected to hold the same distribution as the test data, which are specifically the observational data in this work. For measurement $X$, it is generally subjected to a specific distribution owing to the uncertainty of observations. Here we assume that it is subject to Gaussian distribution $\mathcal{N}(\bar{X},\sigma^2)$, where $\bar{X}$ is the mean of $X$ and $\sigma$ is the corresponding error. However, there are no errors in the measurements simulated by the cosmological model. Therefore, the training sets should be transformed to the same distribution as the observational one before training the ANN. In addition, previous work has shown that adding additional noise to the input data is equivalent to Tikhonov regularization, which could enhance the generalization of trained NNs \citep{Bishop:1995}. Therefore, based on the error level of observational data that is to be used for cosmological parameter estimation, we add Gaussian random noise to the training set to avoid inconsistency of distribution and overfitting, as well as enhance the generalization of the trained model.

At each epoch of the training process, Gaussian noise $\mathcal{N}(0, A^2\sigma^2)$ will be generated and added to each sample of the training set, where $A$ is a coefficient that is $\sim 1$. Note that in the training process of the ANN different noise samples will be generated at each epoch. Therefore, after the epoch of $2\times10^3$, the ANN will be able to statistically learn the distribution of the observational data. In order to reduce the dependence of trained network models on specific experimental observation errors, we take the coefficient $A$ subject to Gaussian distribution $\mathcal{N}(0,0.25)$ that can ensure $|A|\in[0, 1.5]$. This indicates that the ANN will learn the cosmological model with different precision, which means that the trained ANN model can also estimate parameters of the cosmological model when using higher-precision experimental data. Therefore, this may greatly enhance the applicability of this method.

\subsubsection{Data Preprocessing}\label{sec:data_preprocessing}

Previous researches show that the performance of ANN can be influenced by the data-preprocessing techniques \citep{Nawi:2013}. In order to improve the performance and convergence of the ANN, we preprocess the training set before feeding them to the ANN. Specifically, we first divide the cosmological parameters in the training set by their eigenvalues, so that the cosmological parameters become numbers of $\sim 1$. Then, the training set is normalized by using the $Z$-Score normalization technique
\begin{equation}
z = \frac{x-\mu}{\sigma},
\end{equation}
where $\mu$ and $\sigma$ are the mean and standard deviation of the measurement $X$ or the corresponding parameters $P$. This data-preprocessing method can reduce the influence of the order-of-magnitude difference between parameters on the result, so that the ANN can be applied to any cosmological parameters.

\subsubsection{Training Process}\label{sec:training_process}

After two steps of preprocessing via the methods of sections \ref{sec:add_noise} and \ref{sec:data_preprocessing}, the training set can be used to train the ANN. Specifically, the key steps of the training process using ECoPANN are as follows:
\begin{itemize}
	\item[1.] Set initial conditions for cosmological parameters, which are intervals of parameters.
	
	\item[2.] Build a class object for the cosmological model and pass it to ECoPANN, and the training set will be simulated automatically via the method of section \ref{sec:get_training_set} by using the class object.
	
	\item[3.] Pass the errors of the observational data to ECoPANN, and then random noise will be automatically added to the training set via the method of section \ref{sec:add_noise}. Furthermore, the training set will be preprocessed using the method of section \ref{sec:data_preprocessing}.
	
	\item[4.] After the training sets are preprocessed, an ANN model will be built automatically via the method of section \ref{sec:hyperparameters} according to the size of the mock data.
	
	\item[5.] Feed the training set to the ANN model, and the model will be well trained after $2\times10^3$ epochs.
	
	\item[6.] Simulate random samples using the observational data and feed them to the well-trained ANN model, and then a chain of parameters will be produced. Note that the length of the chain is equal to the number of random samples.
	
	\item[7.] Posterior distribution of parameters can be further obtained by using the chain. Then, the parameter space to be learned will be updated according to the posterior distribution of parameters.
	
	\item[8.] Obtain several chains of parameters by repeating steps 2-7. Then, these chains can be used for parameter inference.
\end{itemize}

To obtain a chain of parameters in step 7, we first generate multiple realizations of a data-like sample by drawing the measurement $X$ via the Gaussian distribution $\mathcal{N}(\bar{X},\sigma^2)$. Then, the chain can be obtained by feeding these simulated samples to the well-trained ANN. We note that the initial conditions of cosmological parameters set in step 1 are general ranges of parameters, which means that the true parameters may not be in these ranges. Therefore, the parameter space should be updated in step 7 before training the next ANN. Specifically, we first obtain the best-fit values and errors of parameters from the chain, by using the public code {\it corner}\footnote{\url{https://pypi.org/project/corner/1.0.0/}}, and then the parameter space is updated to $[P-5\sigma_p, P+5\sigma_p]$. Note that the best-fit values here refer to the marginalized means of the posterior distribution.

\subsubsection{Parameter Inference}\label{sec:params_inference}

In the training process, parameters of the ANN ($W$ and $b$ in Equation \ref{eq:def1}) will be updated after each iteration. It should be noted that the parameters of the ANN are initialized randomly before the training process. Thus, given specific hyperparameters and a training set, two initializations of the parameters in the ANN will lead to two different chains of cosmological parameters. To eliminate the effect of the initialization of parameters in the ANN on the results of cosmological parameters, we obtain multiple chains of cosmological parameters by training multiple ANNs in step 8 of section \ref{sec:training_process}, and then we use them to estimate cosmological parameters.

\section{\bf Application to CMB experiments}\label{sec:application_CMB}

To test the capability of the ANN in estimating cosmological parameters, we constrain parameters of the $\Lambda$CDM model using the temperature power spectrum of CMB observations. We will first test the ANN method with the simulated CMB data and then with the Planck CMB observational data.

\subsection{Power Spectrum}\label{sec:prism_power_spectrum}

The mock CMB observations used in our analysis are simulated based on the Polarized Radiation Imaging and Spectroscopy Mission (PRISM; \citet{Andre:2014}), by using the Parameter Forecast for Future CMB Experiments code \citep{Perotto:2006}. The fiducial values of parameters of the $\Lambda$CDM cosmological model are set as follows:
\begin{align}\label{equ:fiducial}
\nonumber H_0&=67.31 \rm ~km ~s^{-1} ~Mpc^{-1}, & \Omega_bh^2 &= 0.02222,\\
\Omega_ch^2 &= 0.1197, & \tau &= 0.078,\\
\nonumber A_s &= 2.19551\times 10^{-9}, & n_s &= 0.9655.
\end{align}
where $H_0$ is the Hubble constant, $\Omega_bh^2$ is the baryon density, $\Omega_ch^2$ is the cold dark matter density, $\tau$ is the optical depth, $A_s$ is the amplitude of primordial inflationary perturbations, and $n_s$ is the spectral index of primordial inflationary perturbations.

\begin{table}
	\centering
	\caption{Experimental Specifications of the PRISM CMB Experiment: Frequency channels, Beam width, Temperature, and Polarization Sensitivities for Each Channel.}\label{tab:prism_specifications}
	\begin{tabular}{c|c|c|c}
		\hline\hline
		Channel & FWHM & $\triangle T$ & $\triangle P$ \\
		(GHz) & (arcmin) & ($\mu$K arcmin) & ($\mu$K arcmin)\\
		\hline
		90  & 5.7 & 3.30 & 4.67 \\
		105 & 4.8 & 2.88 & 4.07 \\
		135 & 3.8 & 2.59 & 3.66 \\
		160 & 3.2 & 2.43 & 3.44 \\
		185 & 2.8 & 2.52 & 3.56 \\
		200 & 2.5 & 2.59 & 3.67 \\
		220 & 2.3 & 2.72 & 3.84 \\
		\hline\hline
	\end{tabular}\\\vspace{5pt}
	{\bf Note.} The sky fraction $f_{\rm sky}=0.8$ for all frequency channels.
\end{table}

The temperature power spectrum is simulated by taking the experimental specifications of PRISM, where the frequency lies within $90-220$GHz. The details of experimental specifications are shown in Table \ref{tab:prism_specifications}, in which the sky fraction $f_{\rm sky}=0.8$ for each frequency channel. We simulate TT, EE, and TE power spectra of CMB, which can be represented as a vector ($C_\ell^{TT}$, $C_\ell^{EE}$, and $C_\ell^{TE}$) with covariance matrix
\begin{equation}
{\rm Cov}_{\ell\ell^\prime}=\frac{2}{(2\ell+1)\triangle\ell f_{\rm sky}}\tilde{C}^2_{\ell}\delta_{\ell\ell^\prime}
\end{equation}
where $\tilde{C}^2_\ell$ runs over $(\tilde{C}_\ell^{TT})^2, ~ (\tilde{C}_\ell^{EE})^2$ and $\frac{1}{2}[(\tilde{C}_\ell^{TE})^2+\tilde{C}_\ell^{TT}\tilde{C}_\ell^{EE}]$ with
\begin{align}
\nonumber \tilde{C}_{\ell}^{TT}&=C_{\ell}^{TT}+N_{\ell}^{TT},\\
\tilde{C}_{\ell}^{EE}&=C_{\ell}^{EE}+N_{\ell}^{EE},\\
\nonumber \tilde{C}_{\ell}^{TE}&=C_{\ell}^{TE}.
\end{align}
Here $N_{\ell}^{TT}$ and $N_{\ell}^{EE}$ are noise power spectra that can be approximated as
\begin{equation}
N_\ell=\theta^2_{\rm FWHM}\sigma_P^2 {\rm exp}\left[\ell(\ell+1)\frac{\theta^2_{\rm FWHM}}{8\ln2}\right]~,
\end{equation}
where $\sigma_P$ is the rms of the instrumental noise, which equals $\triangle T$ for TT power spectrum and $\triangle P$ for EE or TE power spectra.

\subsection{Estimating Parameters}\label{sec:prism_predict_params}

The ANN is trained with the simulated CMB temperature power spectra generated by the Python package of CAMB\footnote{\url{http://camb.readthedocs.io/en/latest}}. Here, we only consider integers that the multipole $\ell\in[30, 2000]$ to train the ANN, so the bin on $\ell$ is 1. The input of the ANN is a temperature power spectrum of CMB, and the outputs are six parameters of the $\Lambda$CDM cosmological model. The specific process of our analysis is unfolded in the following two steps.

\begin{table}
	\centering
	\caption{1$\sigma$ Constraints on Parameters of the $\Lambda$CDM Model Using the Temperature Power Spectrum of the PRISM CMB.}\label{tab:params_prism}
	\begin{tabular}{c|c|c}
		\hline\hline
		& \multicolumn{2}{c}{Methods}  \\
		\cline{2-3}
		Parameters & MCMC & ANN \\
		\hline
		$H_0$         & $67.322\pm0.757$    & $67.293\pm0.784$    \\
		$\Omega_bh^2$ & $0.02228\pm0.00014$ & $0.02222\pm0.00014$ \\
		$\Omega_ch^2$ & $0.11966\pm0.00180$ & $0.11971\pm0.00188$ \\
		$\tau$		  & $0.07799\pm0.01933$	& $0.07895\pm0.02019$ \\
		$10^9A_s$     & $2.19519\pm0.07935$ & $2.19883\pm0.08155$ \\
		$n_s$         & $0.96568\pm0.00411$ & $0.96572\pm0.00425$ \\
		\hline\hline
	\end{tabular}
\end{table}

First, we fit the PRISM CMB temperature power spectrum TT to the $\Lambda$CDM cosmological model using the MCMC method. Here, {\it emcee} \citep{Foreman-Mackey:2013}, a Python module that achieves the MCMC method, is used to constrain the cosmological parameters. During the constraining procedures, 100,000 MCMC chains are generated, and then the best-fit values with 1$\sigma$ errors of these parameters are calculated from the MCMC chains by using {\it corner}, as shown in Table \ref{tab:params_prism}. It is obvious that the best-fit values are consistent with the fiducial ones, and the deviations from the fiducial values are $0.016\sigma$, $0.060\sigma$, $0.024\sigma$, $0.001\sigma$, $0.004\sigma$, and $0.044\sigma$, respectively.

\begin{table}
	\centering
	\caption{Setting of Initial Conditions of Cosmological Parameters in Estimating Cosmological Parameters with the ANN.}\label{tab:initial_condition_TT}
	\begin{tabular}{c|c|c}
		\hline\hline
		Parameters & Minimum & Maximum  \\
		\hline
		$H_0$         & 75     & 80     \\
		$\Omega_bh^2$ & 0.0236 & 0.0250 \\
		$\Omega_ch^2$ & 0.09   & 0.10   \\
		$\tau$		  & 0.28   & 0.35   \\
		$10^9A_s$     & 3.0    & 3.8    \\
		$n_s$         & 1.0    & 1.1    \\
		\hline\hline
	\end{tabular}
\end{table}
\begin{figure*}
	\centering
	\includegraphics[width=0.9\textwidth]{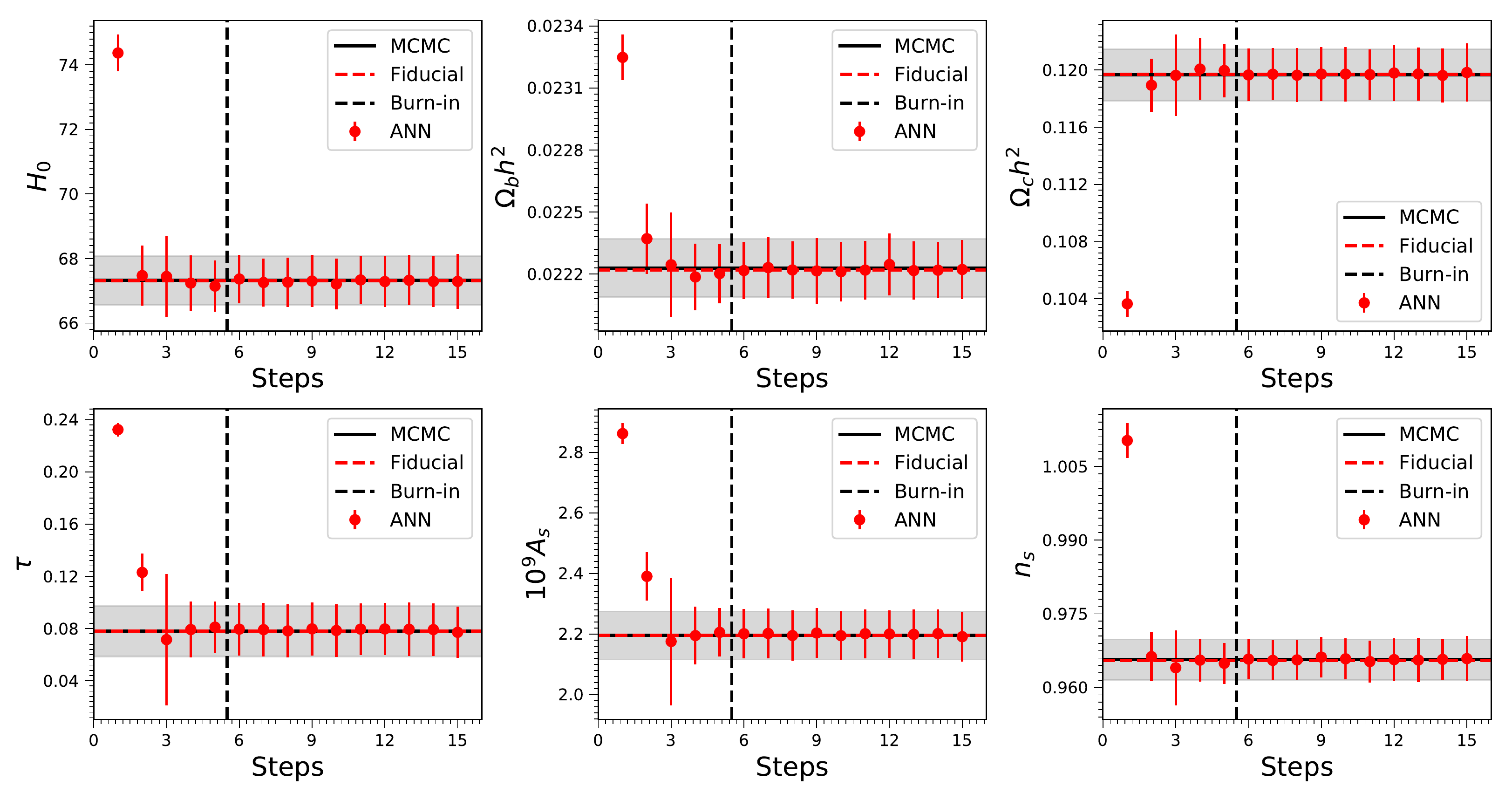}
	\caption{Best-fit values and $1\sigma$ errors of cosmological parameters as a function of steps. The red circles with error bars are the results of the ANN method, the black solid lines and gray areas are those of the MCMC method, and the red dashed lines represent the fiducial values of cosmological parameters.}\label{fig:steps_prism}
\end{figure*}
\begin{figure*}
	\centering
	\includegraphics[width=0.9\textwidth]{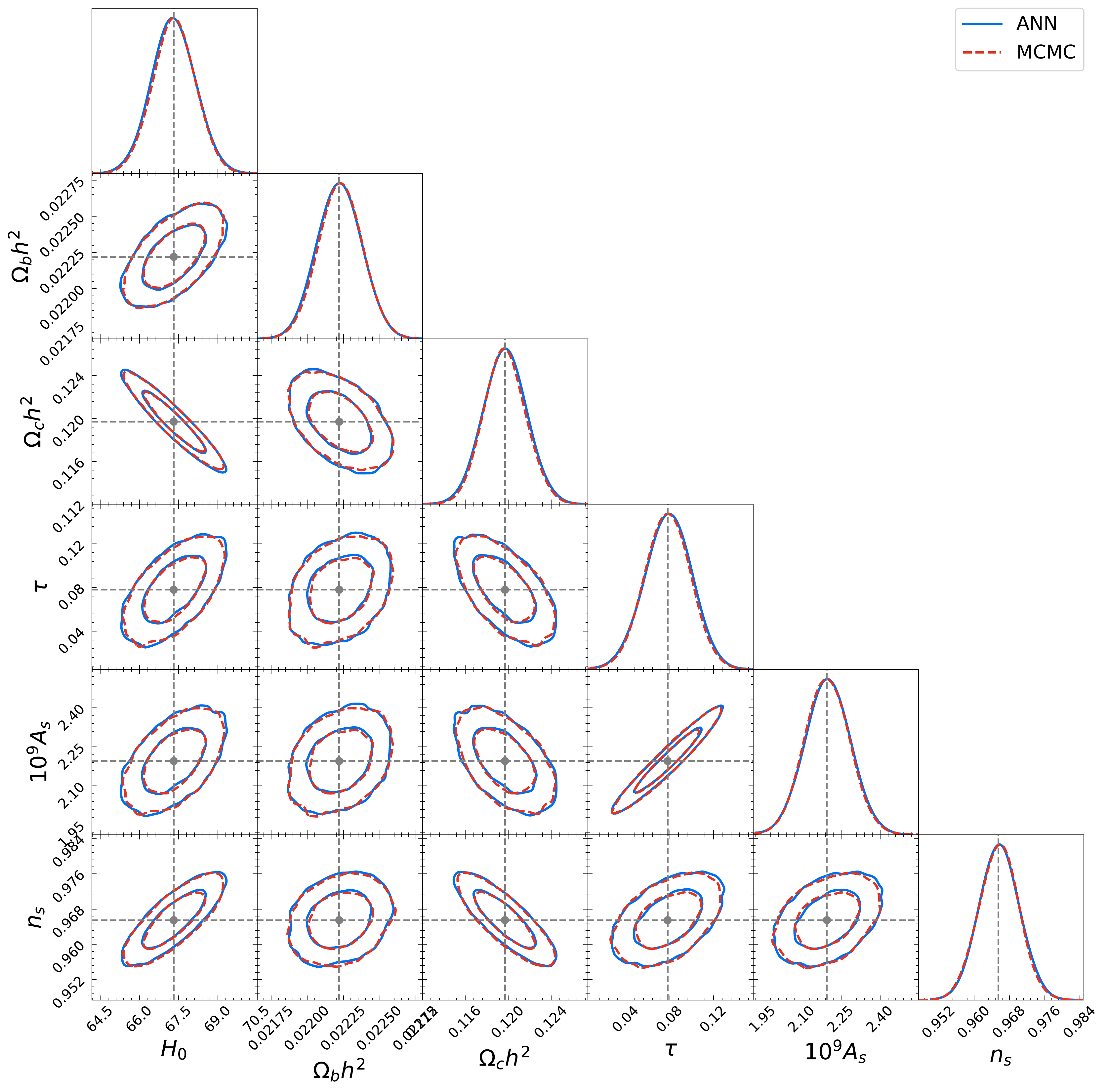}
	\caption{One-dimensional and two-dimensional marginalized distributions with 1$\sigma$ and 2$\sigma$ contours of $H_0$, $\Omega_bh^2$, $\Omega_ch^2$, $\tau$, $A_s$, and $n_s$ constrained from the temperature power spectrum of the PRISM CMB. The blue solid lines are the results of the ANN method, the red dashed lines represent those of the MCMC method, and the gray circles are the fiducial values of the cosmological parameters.}\label{fig:contour_prism}
\end{figure*}

Second, we constrain the cosmological parameters with the ANN, by using the method illustrated in section \ref{sec:training_parameterInference}. We first set initial conditions for the cosmological parameters, as shown in Table \ref{tab:initial_condition_TT}. In order to test the feasibility and reliability of the training strategy of section \ref{sec:training_process}, we set the initial conditions so that they completely deviate from the fiducial values (Equation \ref{equ:fiducial}). The input of the ANN is a spectrum, while the outputs are six cosmological parameters. In the training process, 5000 temperature power spectra are used to train the ANN. Following the training process of section \ref{sec:training_process}, we train 15 ANNs and obtain 15 chains of parameters.

We calculate the best-fit values and $1\sigma$ errors using the 15 chains and draw them in Figure \ref{fig:steps_prism}. The red circles with error bars are the results of the ANN method, while the black solid lines and gray areas represent the best-fit values and $1\sigma$ errors, respectively, of the parameters obtained by the MCMC methods. In this figure, 15 sets of results correspond to 15 steps, which means that an ANN is trained and a chain is obtained in each step. We can see that the results of the ANN deviate greatly from the fiducial values (red dashed lines) at the first step, but as the number of steps increases, both the best-fit values and errors tend to be stable and eventually coincide with the fiducial values, and they also coincide with the results of the MCMC method. This indicates that the ANN method can accurately constrain parameters even if biased initial conditions are given. The reason is that, after training an ANN, the parameter space will be updated according to the posterior distribution of the cosmological parameters, and then the new parameter space will be used to train the next ANN. This shows the feasibility of the training strategy illustrated in section \ref{sec:training_process}.

As shown in Figure \ref{fig:steps_prism}, the results of the cosmological parameters are not stable for the first five steps; thus, the chains in the early part of the steps must be ignored in parameter inference, and we call this part burn-in. Therefore, the ANN chains after the black dashed line are taken in parameter inference. The best-fit values and $1\sigma$ errors obtained from these chains are shown in Table \ref{tab:params_prism}, and we also plot the distributions of the parameters in Figure \ref{fig:contour_prism} (blue solid lines). These results are obviously consistent with the fiducial values of cosmological parameters (gray circles), and they are almost the same as those of the MCMC method (red dashed lines). Furthermore, we can calculate the deviations between the ANN results and the fiducial values according to Table \ref{tab:params_prism}. The deviations of the six cosmological parameters are $0.022\sigma$, $0.009\sigma$, $0.004\sigma$, $0.047\sigma$, $0.041\sigma$, and $0.051\sigma$, respectively, which are quite small. The mean deviation of the six cosmological parameters is $0.029\sigma$, which is similar to that of the MCMC ($0.025\sigma$). In addition, for the errors of the cosmological parameters, the mean relative deviation between the ANN results and the MCMC results is $3.5\%$, which means that the errors of parameters based on the ANN are very similar to those based on the MCMC. Therefore, the ANN method is capable of estimating cosmological parameters with high accuracy.

We note that the length of the burn-in phase is affected by the initial conditions of parameters. In the analysis above, $\sim 10\sigma$ biased initial conditions are selected before training the first ANN model, which lead to the burn-in phase containing five steps. The ANN can make a reasonable prediction for samples whose parameters are located in the parameter space of the training set. Therefore, if good initial conditions are selected to cover the posterior probability distribution of the parameters, the ANN will accurately predict the cosmological parameters in the first step, and thus it would reduce the burn-in phase.

\begin{figure*}
	\centering
	\includegraphics[width=0.9\textwidth]{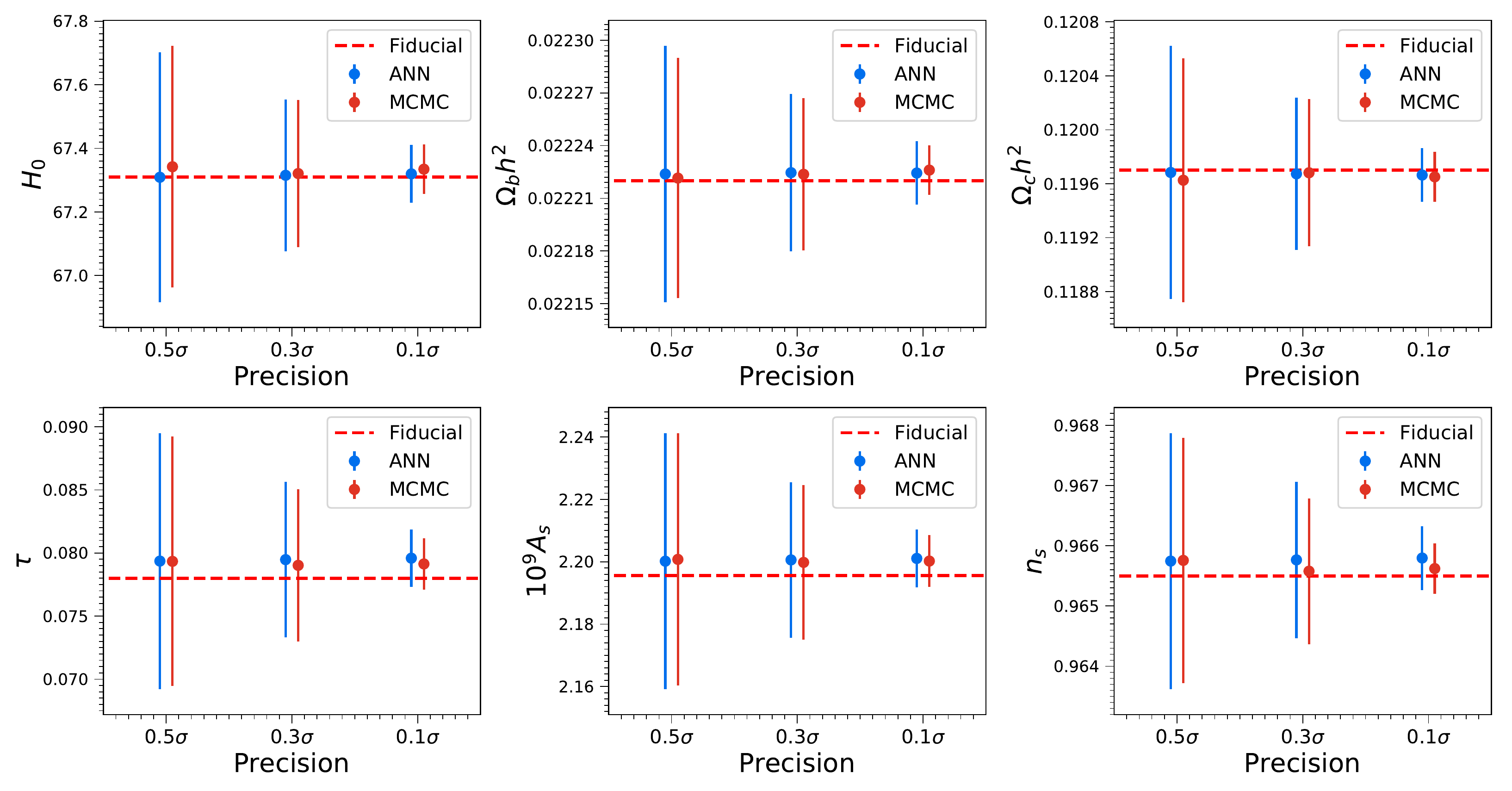}
	\caption{Best-fit values and $1\sigma$ errors of cosmological parameters constrained from higher-precision CMB samples: $0.5\sigma$, $0.3\sigma$, and $0.1\sigma$, respectively, where $\sigma$ is the error of the PRISM CMB. See the text for details.}\label{fig:prism_higher_precision}
\end{figure*}

\subsection{Higher-precision Experiments}\label{sec:test_higher_precision_mission}

As we illustrated in section \ref{sec:add_noise} that different levels of noise are added to the training set. Therefore, in theory, the trained ANNs can be used to estimate cosmological parameters for higher-precision experimental data sets. To test this, we take the well-trained ANNs of section \ref{sec:prism_predict_params} to estimate cosmological parameters using higher-precision CMB samples: $0.5\sigma$, $0.3\sigma$, and $0.1\sigma$, respectively, where $\sigma$ is the error of the PRISM CMB. We call these three samples as sample (a), sample (b), and sample (c), respectively.

The results of the ANN and MCMC methods are shown in Figure \ref{fig:prism_higher_precision}, where the red dashed lines are the fiducial values of the cosmological parameters (Equation \ref{equ:fiducial}). The mean deviations between the ANN results and the fiducial values for the three CMB samples are $0.073\sigma$, $0.137\sigma$, and $0.397\sigma$ respectively. This means that with the improvement of the observational precision, the deviation between the parameters obtained by the ANN and the true values will increase, which is reasonable. For sample (c), the mean deviation is about an order of magnitude larger than that of the PRISM CMB ($0.029\sigma$; see section \ref{sec:prism_predict_params}), which may not be acceptable. However, for samples (a) and (b), the mean deviations are not very large compared to that of the PRISM CMB, which may be acceptable. Furthermore, we can see that the errors of parameters based on the ANN are similar to those based on MCMC. For the errors of the cosmological parameters of the three CMB samples, the mean relative deviations between the ANN results and the MCMC results are $3.7\%$, $3.2\%$, and $17.2\%$, respectively. For samples (a) and (b), these mean relative deviations are similar to that of the PRISM CMB ($3.5\%$; see section \ref{sec:prism_predict_params}), while for sample (c), it is a little larger than that of the PRISM CMB. These results indicate that the ANN trained on the PRISM CMB still performs well in experiments where the precision is increased by about three times. Therefore, the ANNs trained for the PRISM CMB can be used to estimate cosmological parameters for CMB observations that have higher precision. 

It should be noted that when estimating cosmological parameters with the ANN, the training process takes up almost all the time, while very little time (about a few seconds) will be taken for estimating parameters with the well-trained ANN. Therefore, this advantage of the ANN method in estimating cosmological parameters for higher-precision observations will greatly reduce the time of parameter inference, which may be very beneficial to the current and future large-scale sky survey experiments.

\subsection{Reliability of ANN in New Experiments}\label{sec:test_reliability_of_ANN_in_new_experiments}

The analysis of section \ref{sec:test_higher_precision_mission} shows that the ANNs trained with the PRISM CMB can be used for parameter estimation of higher-precision CMB samples, even if the CMB sample has 30\% uncertainties of the PRISM CMB. We note that the samples (a), (b), and (c) used in section \ref{sec:test_higher_precision_mission} have the same fiducial values (Equation \ref{equ:fiducial}) as the PRISM CMB. Moreover, a specific parameter space is learned by an ANN after the training process. Therefore, in theory, a well-trained ANN can only be used to estimate parameters for observations whose true parameters are included in the learned parameter space. This means that for observations whose true parameters exceed the learned parameter space, the ANN should be retrained before estimating parameters.

\begin{table}
	\centering
	\caption{Parameter Space Learned by the ANNs of Section \ref{sec:prism_predict_params} (ANNs after Burn-in in Figure \ref{fig:steps_prism}).}\label{tab:prism_tt_ann_params_space}
	\begin{tabular}{c|c|c|c}
		\hline\hline
		Parameters    & Minimum & Median  & Maximum \\
		\hline
		$H_0$         & 63.401  & 67.278  & 71.155  \\
		$\Omega_bh^2$ & 0.02150 & 0.02222 & 0.02294 \\
		$\Omega_ch^2$ & 0.11037 & 0.11972 & 0.12908 \\
		$\tau$		  & 0.00300 & 0.07895 & 0.18019 \\
		$10^9A_s$     & 1.79385 & 2.20018 & 2.60650 \\
		$n_s$         & 0.94460 & 0.96563 & 0.98664 \\
		\hline\hline
	\end{tabular}
\end{table}

For the ANNs in section \ref{sec:prism_predict_params} (ANNs after burn-in in Figure \ref{fig:steps_prism}), the mean parameter space learned by them is shown in Table \ref{tab:prism_tt_ann_params_space}. As we illustrated in section \ref{sec:get_training_set}, the parameter space to be learned is set to $[P-5\sigma_p, P+5\sigma_p]$, where $P$ is the best-fit value of the posterior distribution of the parameter and $\sigma_p$ is the corresponding $1\sigma$ error, which can be found in Table \ref{tab:params_prism}. We can see that the medians of the parameter space in Table \ref{tab:prism_tt_ann_params_space} are similar to the best-fit values in Table \ref{tab:params_prism}. Note that the optical depth $\tau$ should be a positive value, and the minimum value of it is set to 0.003 to avoid errors in CAMB. Therefore, the parameter space of $\tau$ is cut off by 0.003, and the minimum value of it in the parameter space is 0.003. From the analysis of section \ref{sec:test_higher_precision_mission}, it is difficult to see the ability of a well-trained ANN in estimating parameters for observations whose true parameters deviated from the learned parameter space, such as a different Hubble constant found by new experiments. To test this, in the simulation of the CMB sample, we adopt six different Hubble constant values that deviated from the median of the parameter space (see Table \ref{tab:prism_tt_ann_params_space}) with $1\sigma_p$, $2\sigma_p$, $3\sigma_p$, $4\sigma_p$, $5\sigma_p$, and $6\sigma_p$, respectively. Here, the deviation is defined as follows:
\begin{equation}
\Delta H_0 = \frac{H_0 - H_{0,\rm med}}{\sigma_p},
\end{equation}
where $ H_{0,\rm med}$ is the median of the Hubble constant in the parameter space. Thus, the $H_0$ values are 68.054, 68.829, 69.604, 70.380, 71.155, and 71.931, respectively.

\begin{figure}
	\centering
	\includegraphics[width=0.45\textwidth]{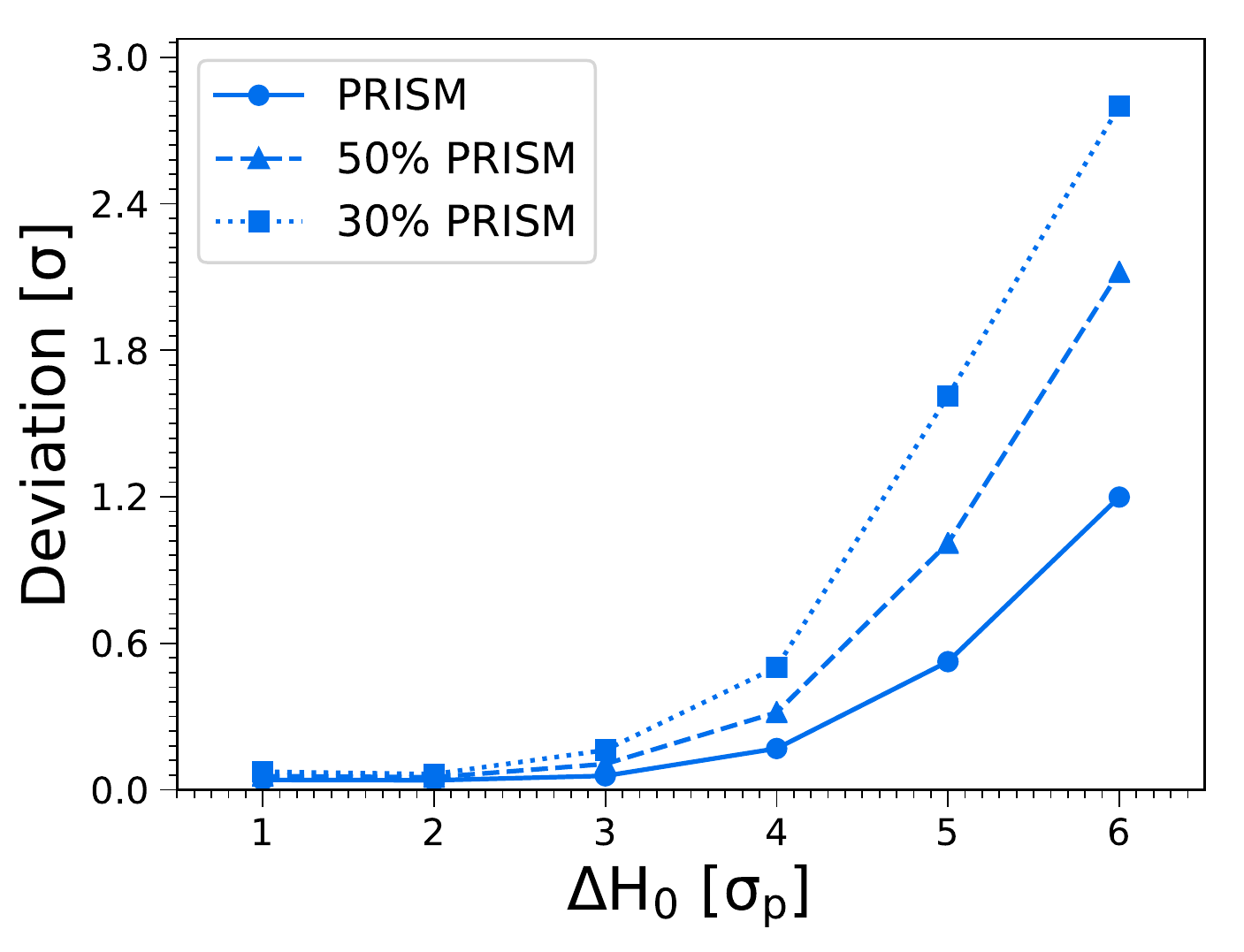}
	\caption{Deviation between the ANN-based $H_0$ and the fiducial value as a function of the deviation of the Hubble constant.}\label{fig:test_biased_H0}
\end{figure}

Using these $H_0$ values, we first simulate six sets of CMB samples based on the experimental specifications of the PRISM experiment (Table \ref{tab:prism_specifications}). Note that for the other five cosmological parameters, the values in Equation \ref{equ:fiducial} are used. Then, we use the well-trained ANNs of section \ref{sec:prism_predict_params} to estimate parameters for these six sets of CMB samples. The deviations between the ANN-based $H_0$ and the fiducial value for the six CMB samples are $0.040\sigma$, $0.039\sigma$, $0.057\sigma$, $0.169\sigma$, $0.525\sigma$, and $1.199\sigma$, respectively, and they are also plotted in Figure \ref{fig:test_biased_H0} (blue solid line). We can see that as the deviation of $H_0$ increases, the ANN result will gradually deviate from the fiducial value. For the deviation of $H_0$ that is $\leq3\sigma_p$, the deviations between the ANN-based $H_0$ and the fiducial value are $\leq0.057\sigma$, which are similar to that of Figure \ref{fig:contour_prism} ($0.029\sigma$). This means that the ANN performs well even if the Hubble constant deviates from the median of the parameter space with $3\sigma_p$. 

Furthermore, we applied the deviated Hubble constant values to samples (a) and (b) in section \ref{sec:test_higher_precision_mission} which have 50\% and 30\% uncertainties of the PRISM CMB, respectively, and another 12 sets of CMB samples are simulated. Then, we use the well-trained ANNs of section \ref{sec:prism_predict_params} and the MCMC method to estimate parameters for these 12 sets of CMB samples. The deviations between the ANN-based $H_0$ and the fiducial value are shown in Figure \ref{fig:test_biased_H0} with the blue dashed line and the blue dotted line. For the deviation of $H_0$ that is $\leq3\sigma_p$, the deviations between the ANN-based $H_0$ and the fiducial value are $\leq0.106\sigma$ for samples with 50\% uncertainties and $\leq0.162\sigma$ for samples with 30\% uncertainties. We can see that all these results are similar to those in section \ref{sec:test_higher_precision_mission}. Therefore, it is reliable to use the ANN trained based on the current observational data to estimate parameters for future higher-precision observations.

\begin{figure}
	\centering
	\includegraphics[width=0.45\textwidth]{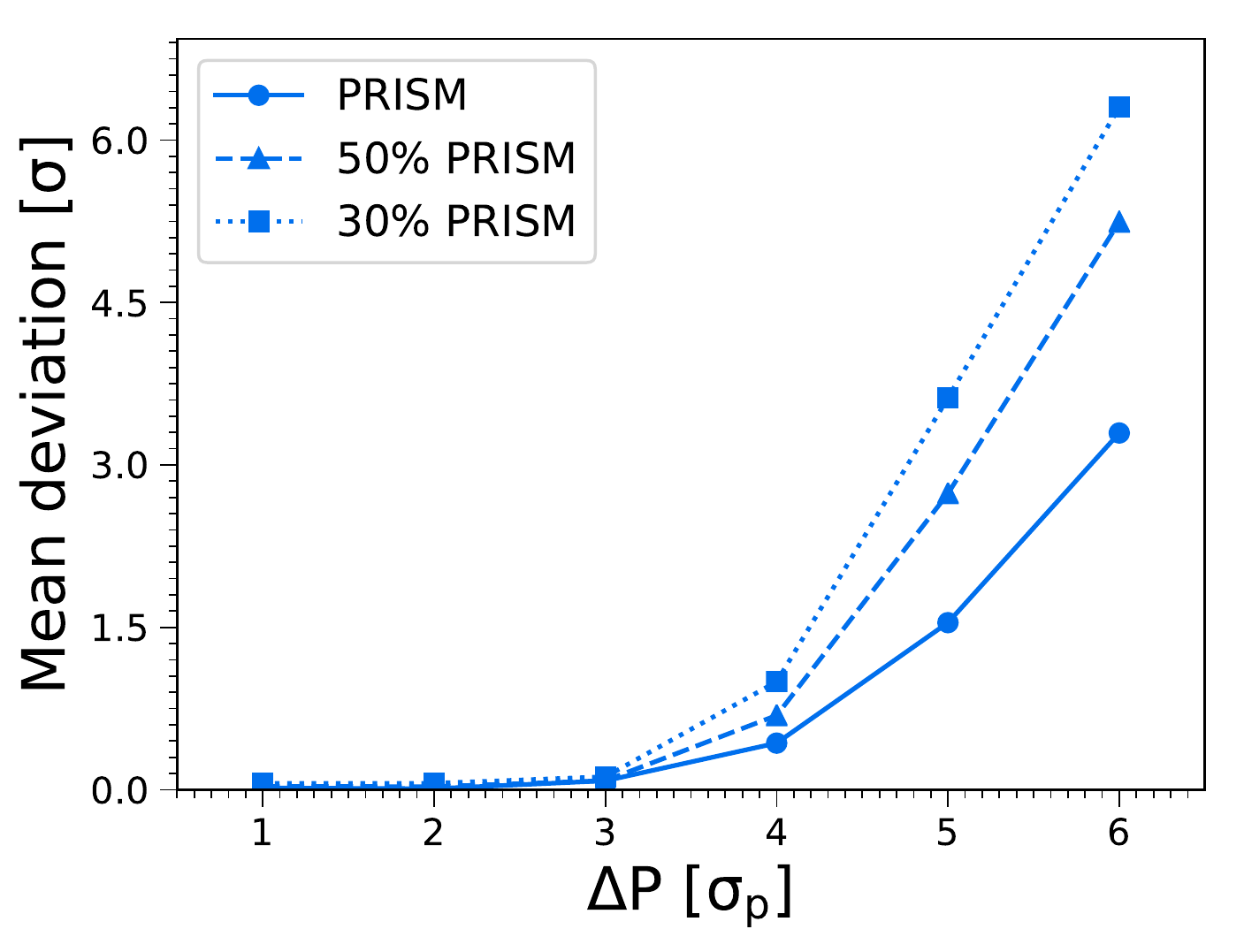}
	\caption{Mean deviations between the six ANN-based cosmological parameters and the fiducial values as a function of the deviation of the parameters.}\label{fig:test_biased_all_params}
\end{figure}

It should be noted that all six cosmological parameters measured by new experiments may have deviations from the learned parameter space. Thus, with the same procedure, we adopt another six different sets of cosmological parameters such that all of them deviated from the median of the parameter space with $1\sigma_p$, $2\sigma_p$, $3\sigma_p$, $4\sigma_p$, $5\sigma_p$, and $6\sigma_p$, respectively. Then, we simulate the CMB samples and use the well-trained ANNs of section \ref{sec:prism_predict_params} to estimate the corresponding cosmological parameters. The mean deviations between the ANN results and the fiducial values are shown in Figure \ref{fig:test_biased_all_params}. We can see that the mean deviation will increase significantly after $3\sigma_p$, which is similar to those of Figure \ref{fig:test_biased_H0}. When the deviation of the cosmological parameter from the median of the parameter space is less than $3\sigma_p$, the deviations between the ANN results and the fiducial values are $\leq0.083\sigma$ for the PRISM CMB samples, and $\leq0.084\sigma$ for the samples with 50\% uncertainties of the PRISM CMB, and $\leq0.120\sigma$ for the samples with 30\% uncertainties of the PRISM CMB. These values can be acceptable in parameter estimations. Therefore, when all the parameters deviate from the median of the parameter space with $3\sigma_p$, the well-trained ANNs can also estimate the cosmological parameters with high accuracy.

As shown in Figures \ref{fig:test_biased_H0} and \ref{fig:test_biased_all_params}, even if the true cosmological parameter deviates from the median of the parameter space with $4\sigma_p$, the ANN results are consistent with the fiducial values within a $1\sigma$ confidence level. However, we note that when the parameter deviates from the median of the parameter space with more than $3\sigma_p$, the ANN results will gradually deviate from the true values. Therefore, when using ECoPANN to estimate parameters, $3\sigma_p$ is taken as a threshold to determine whether ANN can estimate the parameters with high accuracy. This means that if the estimated best-fit values of the parameters are not included in the range of $[P-3\sigma_p, P+3\sigma_p]$, the ANN should be retrained for the new experiments. This advantage of the ANN may be very helpful for parameter estimation of some sky survey experiments.

\subsection{Test with Planck CMB}\label{sec:planck_predict_params}

\begin{table}
	\centering
	\caption{1$\sigma$ Constraints on Parameters of the $\Lambda$CDM Model Using the Temperature Power Spectrum of the Planck CMB.}\label{tab:params_planck}
	\begin{tabular}{c|c|c}
		\hline\hline
		& \multicolumn{2}{c}{Methods} \\
		\cline{2-3}
		Parameters & MCMC & ANN \\
		\hline
		$H_0$         & $67.918\pm1.197$    & $67.918\pm1.240$     \\
		$\Omega_bh^2$ & $0.02237\pm0.00023$ & $0.02237\pm0.00024$  \\
		$\Omega_ch^2$ & $0.11854\pm0.00275$ & $0.11847\pm0.00279$  \\
		$\tau$		  & $0.12913\pm0.03263$ & $0.12993\pm0.03170$  \\
		$10^9A_s$     & $2.42592\pm0.14963$ & $2.42893\pm0.14252$  \\
		$n_s$         & $0.96841\pm0.00680$ & $0.96865\pm0.00713$  \\
		\hline\hline
	\end{tabular}
\end{table}

\begin{figure*}
	\centering
	\includegraphics[width=0.9\textwidth]{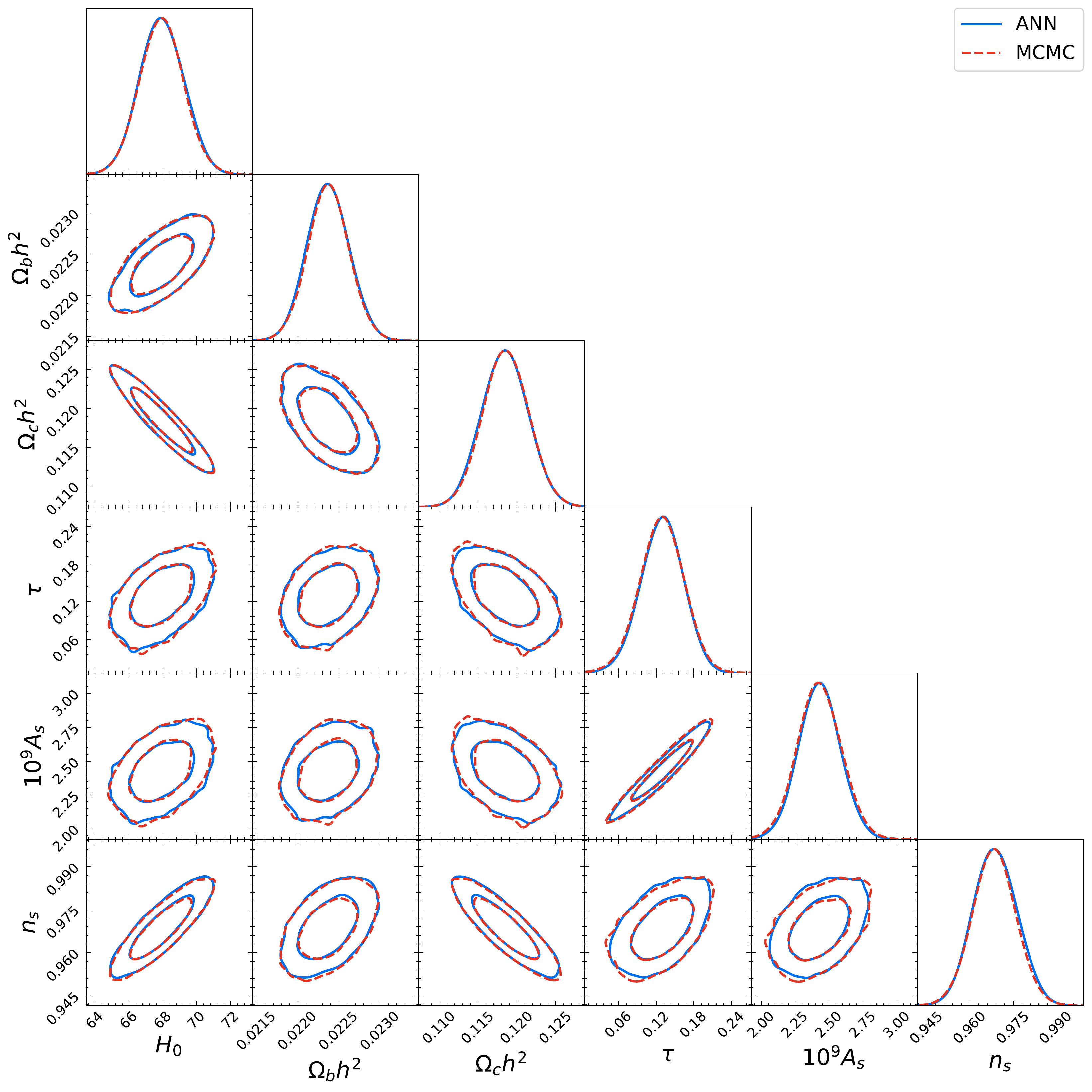}
	\caption{Same as Figure \ref{fig:contour_prism}, except now using the Planck2015 temperature power spectrum.}\label{fig:contour_planck2015}
\end{figure*}

In the analysis of section \ref{sec:prism_predict_params}, the ANN is capable of estimating the cosmological parameters with high accuracy for the simulated CMB observation. In theory, this pipeline can also be used for the observational CMB missions. Following the same procedures, we estimate the cosmological parameters of the $\Lambda$CDM model with Planck2015 temperature power spectrum {\it COM\_PowerSpect\_CMB\_R2.02.fits}\footnote{\url{http://pla.esac.esa.int/pla/\#cosmology}}. We use the MCMC and ANN methods simultaneously to estimate the six cosmological parameters. The results of these two methods are listed in Table \ref{tab:params_planck}, and the one-dimensional and two-dimensional marginalized distributions of the cosmological parameters are shown in Figure \ref{fig:contour_planck2015}, in which the blue solid lines represent the results of the ANN method, while the red dashed lines are those of the MCMC method. Obviously, the results of these two methods are consistent with each other. More specifically, the deviations between the ANN results and the MCMC results for the six cosmological parameters are $0.000\sigma$, $0.016\sigma$, $0.018\sigma$, $0.018\sigma$, $0.015\sigma$, and $0.025\sigma$, respectively. This indicates that the ANN method can almost get the same results as the MCMC method. Therefore, our method can also be used for parameter estimation of observational data, which means it has a wide range of applicability.

\section{\bf Joint constraint on parameters}\label{sec:joint_constraint}

The analysis of section \ref{sec:application_CMB} shows that the ANN method performs very well in estimating cosmological parameters with one dataset. However, multiple data sets from different experiments are usually required to simultaneously constrain cosmological parameters, which is not possible for the ANN model of Figure \ref{fig:nn_model}. To do this, we expand the ANN model of Figure \ref{fig:nn_model} to a multibranch network to achieve a joint constraint on parameters. In this section, we will first illustrate the multibranch network and then test it using the simulated CMB, SN Ia, and BAO data sets.

\subsection{Multibranch Network}

\begin{figure}
	\centering
	\includegraphics[width=0.45\textwidth, angle=0]{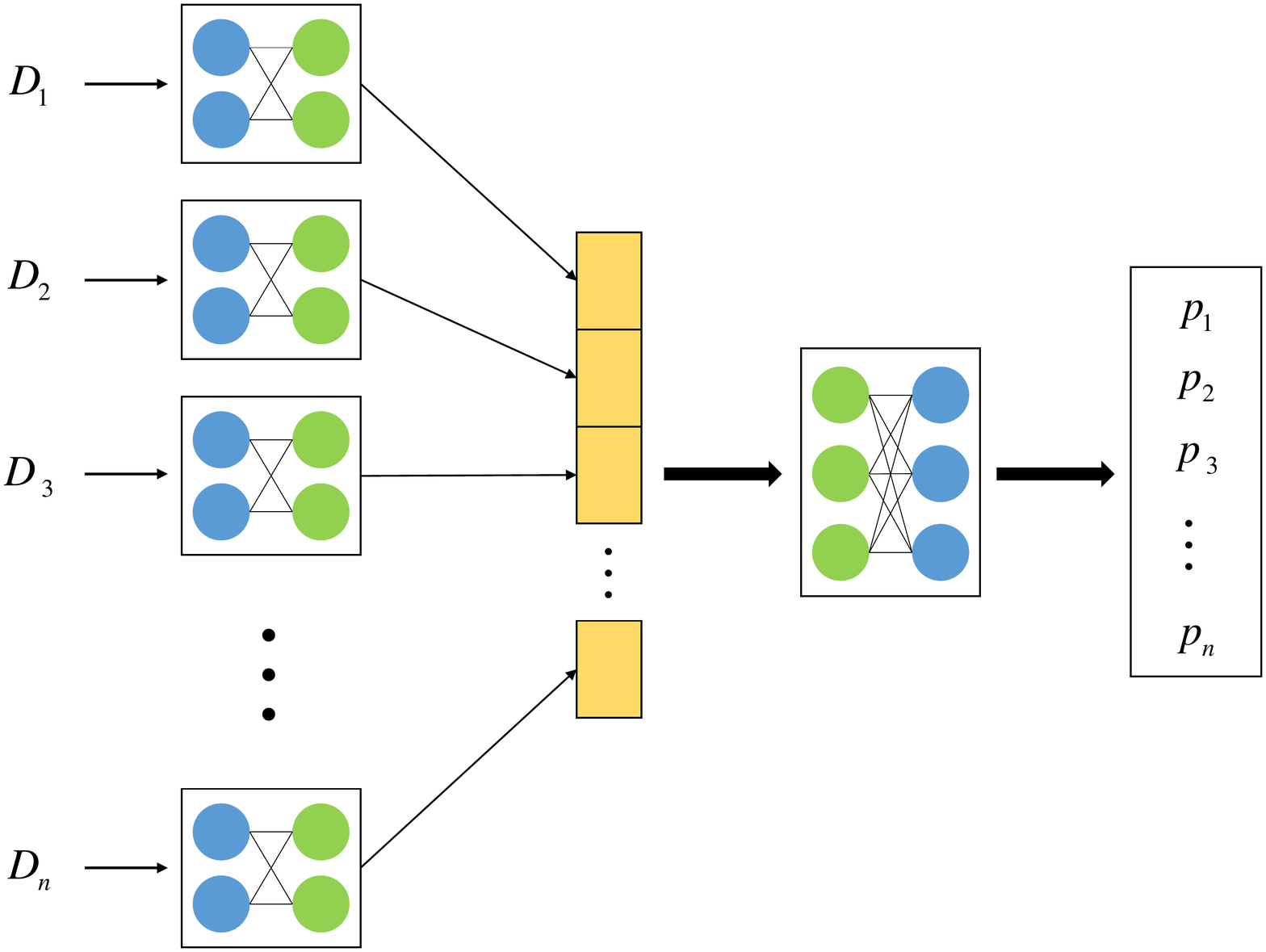}
	\caption{General structure of a multibranch network that achieves a joint constraint on cosmological parameters. The inputs are multiple data sets $\{D_1, D_2, ..., D_n\}$ from different experiments. The outputs are parameters of the cosmological model to be estimated.}\label{fig:multi_branch_net}
\end{figure}

The general structure of a multibranch network is shown in Figure \ref{fig:multi_branch_net}, where the inputs are multiple data sets from different experiments and the outputs are the cosmological parameters to be estimated. Each branch accepts one component of the observational data sets and processes them independently in shallow layers. Then, intermediate features are concatenated and fed into the remaining part of the network to obtain the estimation of parameters. In our network structure, each branch consists of four fully connected layers while the remaining part has two.

To obtain a well-behaved joint multibranch estimator, we firstly train an independent ANN for every component of the data sets and copy the first four layers to the corresponding branch, which will effectively extract features of the observational data and accelerate the training of the ANN. Then, we keep the weights of the branches and optimize the remaining part of the network by back-propagation. Finally, we fine-tune the entire network. Besides, the parameters of the trained ANN can be used as the initialization of the ANN in the next step, which can also effectively improve the training speed of the network.

\subsection{Test with the CMB, SNe Ia, and BAOs}

In this section, we test the multibranch network by constraining six cosmological parameters of the $\Lambda$CDM model with the simulated CMB, SN Ia, and BAO data sets. Similarly, we achieve our analysis by comparing the results of the ANN method with those of the MCMC method.

\subsubsection{Data Simulations}

Taking the parameters of Equation \ref{equ:fiducial} as the fiducial cosmology, we simulate the CMB observation based on the experimental specifications of the PRISM experiment (Table \ref{tab:prism_specifications}), the SNe Ia based on the future Wide-Field Infra-Red Survey Telescope (WFIRST) experiment \citep{Spergel:2015}, and the BAO measurements based on the future SKA2 survey \citep{Bull:2015}. In addition to the temperature power spectrum, the expanded pipeline also involves the polarization power spectrum of the PRISM CMB. The total number of SNe Ia predicted by WFIRST is 2725, which is expected in each $\triangle z=0.1$ bin for redshift in the range of $0.1<z<1.7$. The photometric measurement error per supernova is $\sigma_{\rm meas} = 0.08$ mag, and the intrinsic dispersion in luminosities is assumed as $\sigma_{\rm int} = 0.09$ mag. The other contribution to statistical errors is gravitational lensing magnification, which is modeled as $\sigma_{\rm lens} = 0.07 \times z$ mag. 

The Square Kilometer Array (SKA) project is an international collaboration to build the world's largest radio telescope, the construction of which is divided into two phases: SKA Phase 1 (SKA1) and SKA Phase 2 (SKA2). SKA2 will achieve an RMS flux sensitivity of $S_{\rm rms} \approx 5 ~\mu\rm Jy$ for a 10,000 hr survey over 30,000 deg$^2$. The expected yield for such a survey is $\sim 10^9$ galaxies between $0.18 < z < 1.84$. These make it powerful in measuring BAOs. Here we take $17$ BAO measurements from \citet{Bull:2015} to estimate parameters. The measurements of BAOs are the Hubble parameter $H(z)$ and the angular diameter distance $D_A(z)$. For the flat $\Lambda$CDM model
\begin{equation}
H(z) = H_0 \sqrt{\Omega_{\rm m}(1+z)^3 + 1-\Omega_{\rm m}}~,
\end{equation}
where $\Omega_{\rm m}=(100^2\Omega_bh^2)/H_0^2+(100^2\Omega_ch^2)/H_0^2$. The luminosity distance is
\begin{equation}
D_L(z) = c\cdot(1+z)\int_{0}^{z}\frac{dz'}{H(z')} ~,
\end{equation}
where $c$ is the speed of light. So, $D_A(z)$ can be calculated using the cosmic distance duality $D_A(z) = D_L/(1+z)^2$. Thus, the BAO measurement is sensitive to $H_0$, $\Omega_bh^2$, and $\Omega_ch^2$. For SNe Ia, the distance modulus
\begin{equation}
\mu(z)=5\log\frac{D_L}{\rm Mpc}+25 ~,
\end{equation}
which is also sensitive to $\Omega_bh^2$ and $\Omega_ch^2$, is usually used to estimate parameters.

\begin{table}
	\centering
	\caption{1$\sigma$ Constraints on Parameters of the $\Lambda$CDM Model Using the Simulated PRISM CMB, SN Ia, and BAO Data Sets.}\label{tab:params_prismSNeBAO}
	\begin{tabular}{c|c|c}
		\hline\hline
		& \multicolumn{2}{c}{Methods} \\
		\cline{2-3}
		Parameters & MCMC & ANN \\
		\hline
		$H_0$         & $67.306\pm0.103$    & $67.305\pm0.103$    \\
		$\Omega_bh^2$ & $0.02223\pm0.00003$ & $0.02223\pm0.00003$ \\
		$\Omega_ch^2$ & $0.11972\pm0.00029$ & $0.11973\pm0.00028$ \\
		$\tau$		  & $0.07828\pm0.00221$ & $0.07813\pm0.00218$ \\
		$10^9A_s$     & $2.19679\pm0.00910$ & $2.19619\pm0.00907$ \\
		$n_s$         & $0.96560\pm0.00135$ & $0.96556\pm0.00133$ \\
		\hline\hline
	\end{tabular}
	\\	
	\textbf{Notes.} The SN Ia data are simulated based on the WFIRST experiment, and the BAO measurements are simulated based on the SKA2 survey.
\end{table}

\begin{figure*}
	\centering
	\includegraphics[width=0.9\textwidth]{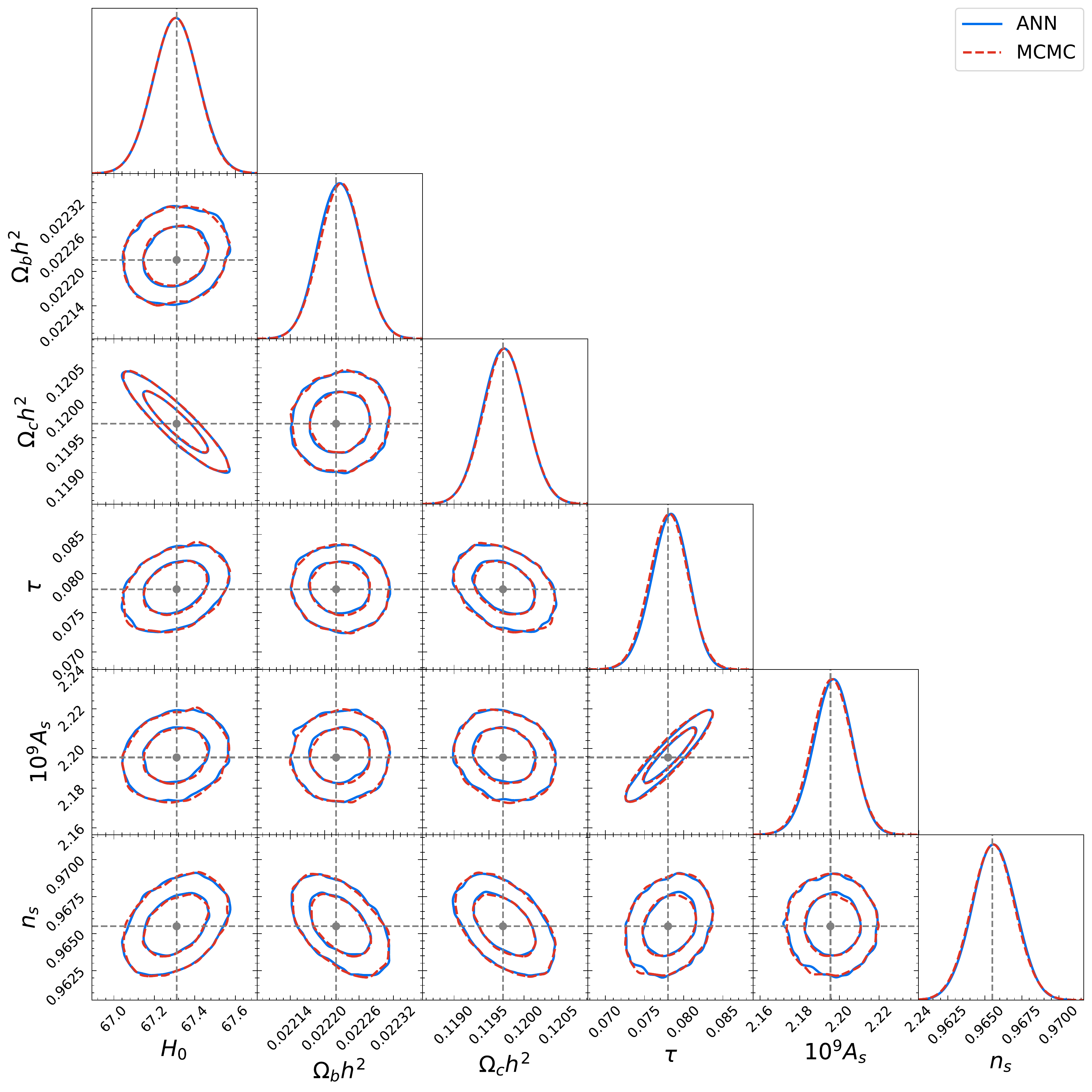}
	\caption{One-dimensional and two-dimensional marginalized distributions with 1$\sigma$ and 2$\sigma$ contours of $H_0$, $\Omega_bh^2$, $\Omega_ch^2$, $\tau$, $A_s$, and $n_s$ constrained from PRISM CMB temperature and polarization spectra, SNe Ia of future WFIRST experiment, and BAO measurements of a future SKA2 survey. The blue solid lines are the results of the ANN method, the red dashed lines represent those of the MCMC method, and the gray circles are the fiducial values of the cosmological parameters.}\label{fig:contour_prismSNeBAO}
\end{figure*}

\subsubsection{Results}

We estimate the cosmological parameters with both the ANN and MCMC methods. The results of the MCMC method are shown in Table \ref{tab:params_prismSNeBAO}, which are consistent with the fiducial cosmological model (Equation \ref{equ:fiducial}) within a $1\sigma$ confidence level, and the deviations from the fiducial values are $0.052\sigma$, $0.239\sigma$, $0.092\sigma$, $0.061\sigma$, $0.075\sigma$, and $0.044\sigma$, respectively. With the same procedure as in section \ref{sec:training_process}, we train the multibranch network with the simulated CMB, SN Ia, and BAO data sets. The inputs consist of six components: TT, EE, and TE spectra of CMB, the distance modulus $\mu(z)$ of SNe Ia, and the Hubble measurements $H(z)$ and the angular diameter distance $D_A(z)$ of BAOs, while the outputs are six cosmological parameters.

After training the ANNs, we obtain six chains that can be used to estimate the cosmological parameters. Finally, we calculate the best-fit values and $1\sigma$ errors by using these chains, shown in Table \ref{tab:params_prismSNeBAO}. Furthermore, we plot the one-dimensional and two-dimensional marginalized distributions with $1\sigma$ and $2\sigma$ contours of the parameters in Figure \ref{fig:contour_prismSNeBAO}, where the blue solid lines represent the results of the ANN method and the red dashed lines are for those of the MCMC method. Obviously, we can see that the results of the ANN method are consistent with the fiducial values (Equation \ref{equ:fiducial}) and are almost the same as the results of the MCMC method. 

In addition, we calculate the deviations between the ANN results and the fiducial values according to Table \ref{tab:params_prismSNeBAO}, which are $0.039\sigma$, $0.189\sigma$, $0.071\sigma$, $0.128\sigma$, $0.141\sigma$, and $0.077\sigma$, respectively. Here, the mean deviation of the six parameters is $0.107\sigma$, which is similar to that of the MCMC ($0.094\sigma$). Moreover, for the errors of the cosmological parameters, the mean relative deviation between the ANN results and the MCMC results is $1.4\%$, which is small enough to be acceptable in parameter estimations. Therefore, the multibranch network is capable of constraining cosmological parameters with high accuracy.

\section{\bf Effect of hyperparameters}\label{sec:test_hyperparameters}

Hyperparameters of the ANN are set to specific values in the upper analysis (see section \ref{sec:hyperparameters}). However, the performance of the ANN may be influenced by the setting of hyperparameters. Thus, in this section we test the effect of hyperparameters on the results of parameter estimations. Specifically, we test the effect of the number of hidden layers, the activation functions, the number of the training sets, and the number of epochs on the results, by using the simulated temperature power spectrum of the PRISM CMB.

\begin{figure*}
	\centering
	\includegraphics[width=0.45\textwidth]{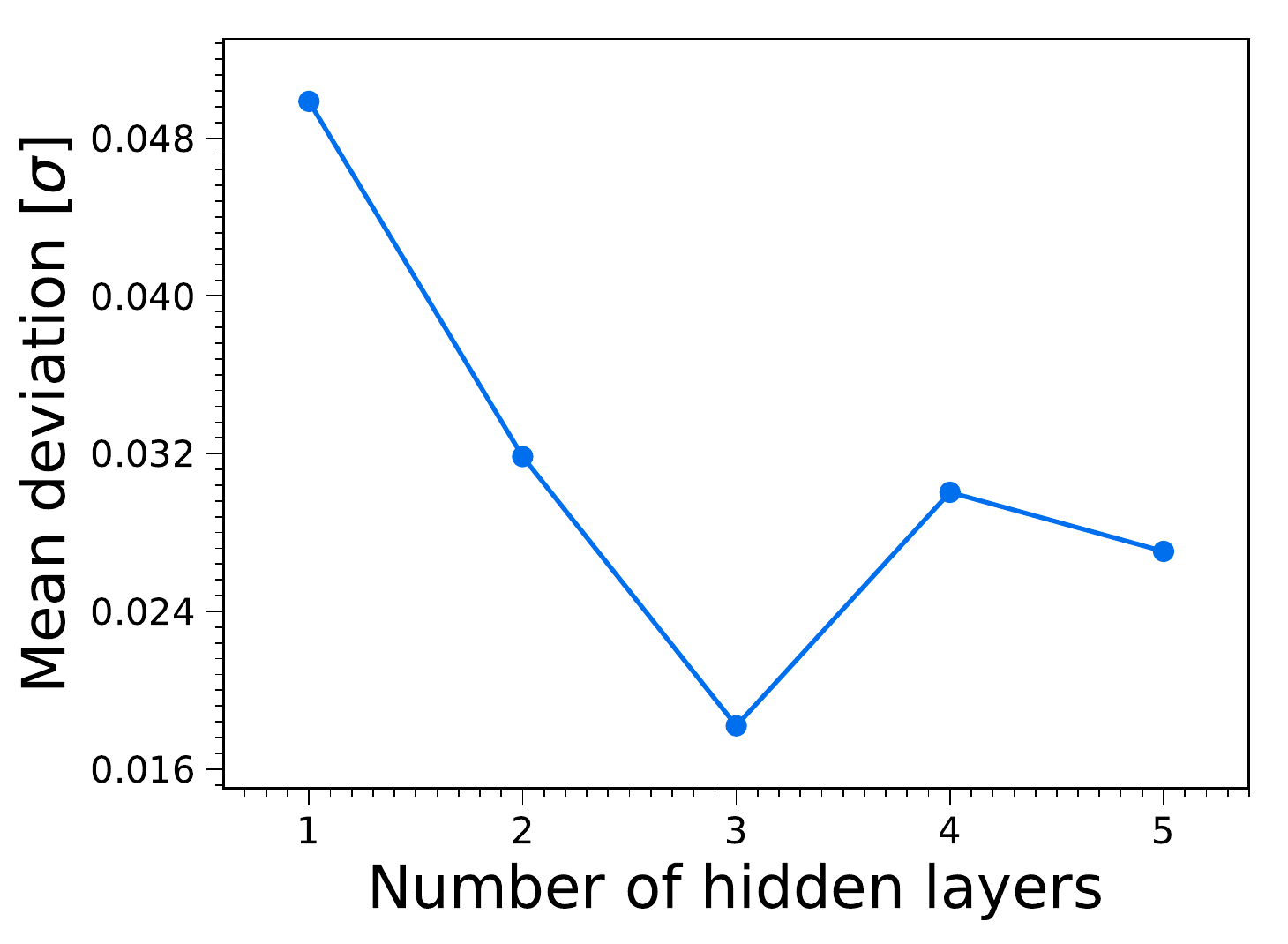}
	\includegraphics[width=0.45\textwidth]{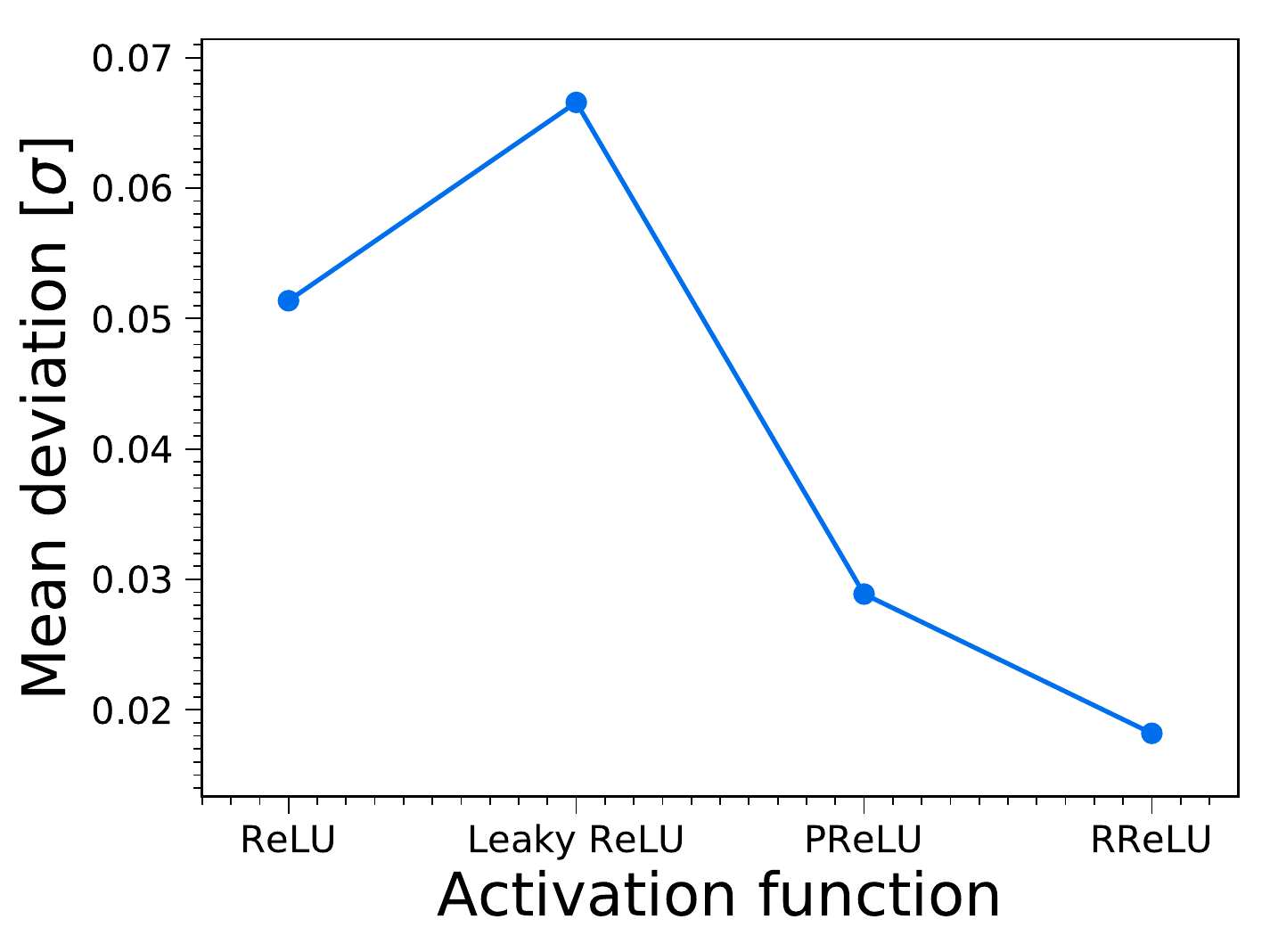}
	\includegraphics[width=0.45\textwidth]{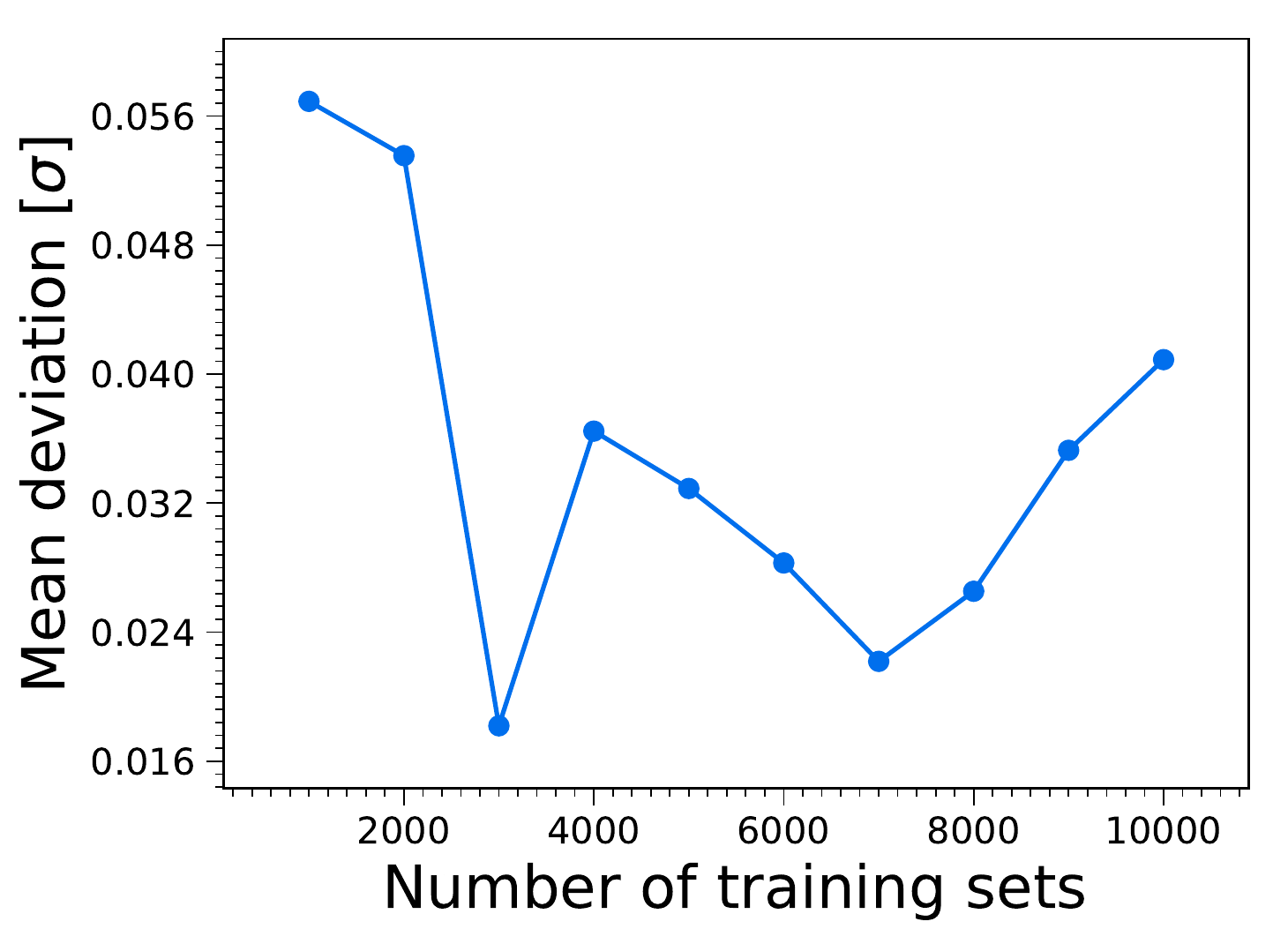}
	\includegraphics[width=0.45\textwidth]{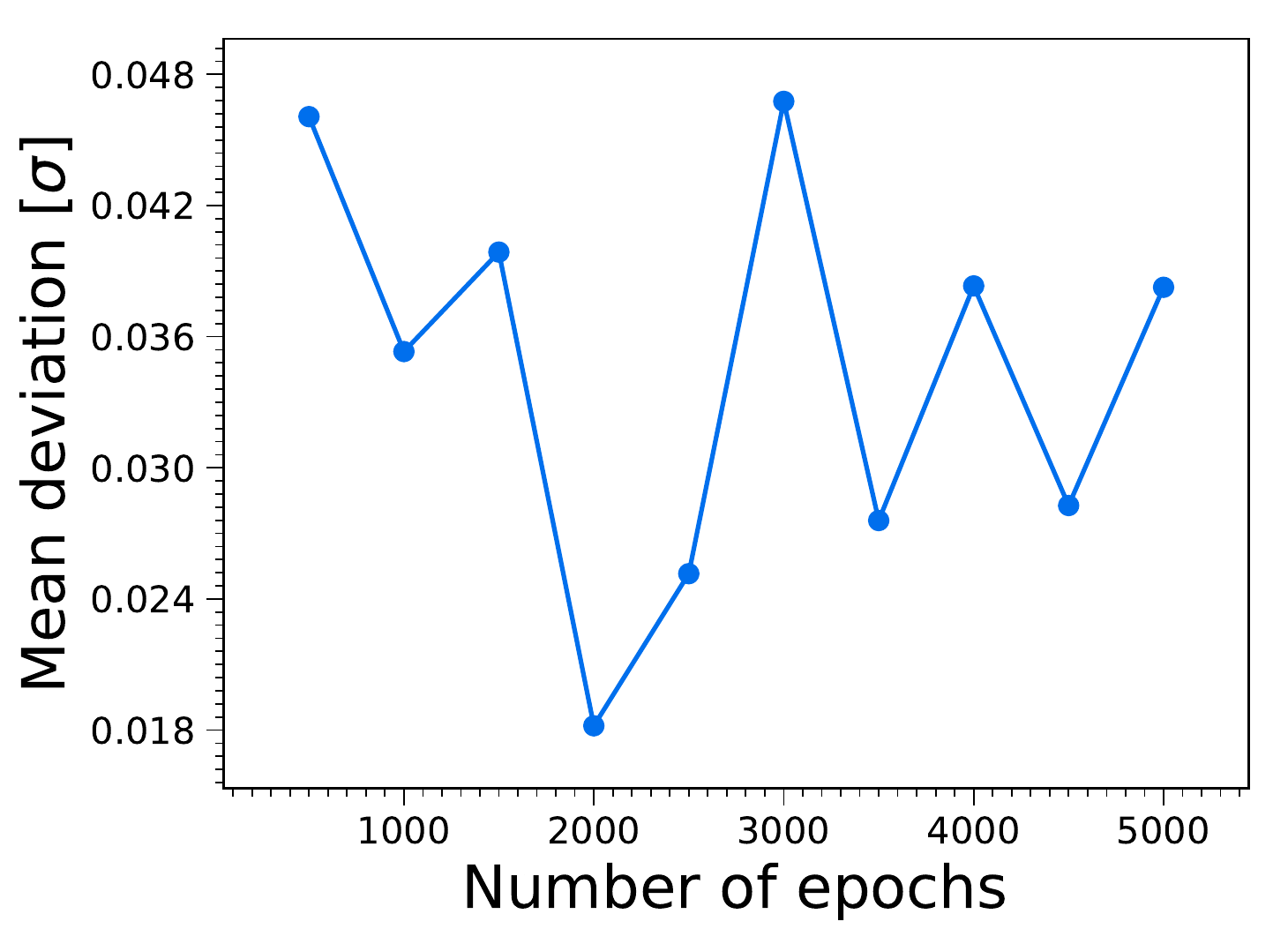}
	\caption{Mean deviations between the ANN results and the fiducial values as a function of the number of hidden layers, the activation function, the number of training sets, and the number of epochs, respectively.}\label{fig:test_hyperparameters}
\end{figure*}

\subsection{The Number of Hidden Layers}

We first test the effect of the number of hidden layers in the ANN model of Figure \ref{fig:nn_model}. We design five different ANN structures with the number of hidden layers from 1 to 5, where the number of neurons in each layer is set according to Equation \ref{equ:neuron_num}. In addition, the activation function is RReLU, and the number of samples in the training set is 3000. Then, five sets of ANNs are trained, and the corresponding chains are obtained according to the procedure of sections \ref{sec:training_process} and \ref{sec:params_inference}. Finally, the best-fit values and $1\sigma$ errors of cosmological parameters can be calculated from these chains.

Furthermore, we calculate the mean deviations between the ANN results and the fiducial ones (Equation \ref{equ:fiducial}). The mean deviation as a function of the number of hidden layers is shown in the top left panel of Figure \ref{fig:test_hyperparameters}, where the maximum deviation is $0.050\sigma$ (for the ANN with one hidden layer) and the minimum deviation is $0.018\sigma$ (for the ANN with three hidden layers). The deviation of $0.050\sigma$ may be acceptable in parameter estimations; thus, this may indicate that the ANN can be used to estimate parameters even when it has one hidden layer. However, in the five structures of the ANN, the structure with three hidden layers has the minimum deviation. Therefore, we adopt the ANN structure that has three hidden layers in our analysis.

\subsection{Activation Function}

To test the effect of activation function on the results of parameter estimations, we select four kinds of rectified units: rectified linear (ReLU), leaky rectified linear (Leaky ReLU), parametric rectified linear (PReLU), and the RReLU activation function used in the analysis above. The ReLU activation function is first used by \citet{Nair:2010}, which is defined as
\begin{equation}\label{equ:relu}
f(x)=\left\{\begin{matrix}
x & \text{if } x \geq 0 \\
0 & \text{if } x < 0 .
\end{matrix}\right.
\end{equation}
The leaky ReLU is introduced by \citet{Maas:2013}, with the mathematical form
\begin{equation}\label{equ:lrelu}
f(x)=\left\{\begin{matrix}
x & \text{if } x \geq 0 \\
\frac{x}{a} & \text{if } x < 0,
\end{matrix}\right.
\end{equation}
where $a$ is a fixed parameter in the range $(1, +\infty)$. In the analysis of \citet{Maas:2013}, the authors suggest setting $a$ to a large number like 100; thus, we set $a$ to be 100 in our analysis. For the PReLU activation function, it is proposed by \citet{prelu}, which has the same mathematical form as the leaky ReLU (Equation \ref{equ:lrelu}). However, $a$ is a learnable parameter to be learned in the training process via back-propagation.

In our analysis, the structure of the ANN with three hidden layers is adopted, and the training set contains 3000 samples. With the same procedure as section \ref{sec:training_process}, we estimate cosmological parameters with ANNs by adopting these four different activation functions. After obtaining chains of parameters, the mean deviations of parameters between the ANN results and fiducial values are calculated, shown in the top right panel of Figure \ref{fig:test_hyperparameters}. The results show that the activation function will affect the performance of the ANN in parameter estimation. In the four activation functions, the superiority of RReLU is more significant than that of the other three activation functions. Therefore, the RReLU activation function is recommended in the task of parameter estimation.

\subsection{The Number of Training Sets}

Previous researches show that the number of training sets also affects the performance of the ANN. To test this, we train the ANN with training sets that have different numbers of samples. Specifically, the number of samples of the training set varies from 1000 to 10,000. In the analysis, an ANN with three hidden layers is adopted, and the activation function is RReLU. With the same procedure, we train ANNs and then obtain the corresponding chains of parameters. Finally, the mean deviations are calculated and are shown in the bottom left panel of Figure \ref{fig:test_hyperparameters}. We can see that for the training set that has 1000 or 2000 samples the deviation is a little larger. However, when the number of training set is more than 3000, the deviation will be relatively lower. It should be noted that the time of training an ANN is related to the amount of data in the training set. Therefore, considering the performance of the ANN and the training time, the number of training sets should be selected reasonably.

\subsection{The Number of Epochs}

The number of epochs may also affect the performance of the ANN. To test this, we train the ANN with a different number of epochs, which varies from 500 to 5000. For other hyperparameters, the ANN with three hidden layers is adopted, the activation function is set to RReLU, and the number of training sets is 3000. With the same procedure, we train ANNs and obtain the corresponding chains of parameters. Finally, we calculate the mean deviations and plot them in the bottom right panel of Figure \ref{fig:test_hyperparameters}. We can see that as the number of epochs increases, the mean deviation will first oscillate violently, and then the oscillation will gradually decrease, and all the deviations are small enough to be acceptable. It should be noted that the time of training an ANN is also related to the number of epochs. Therefore, the number of epochs should be selected reasonably.

\section{\bf Discussions}\label{sec:discussion}

\subsection{Initial Conditions of Parameters}\label{sec:discussion_initial_conditions}

\begin{table}
	\centering
	\caption{Same as Table \ref{tab:initial_condition_TT}, but Now the Initial Conditions Are Based on the Planck2015 Results \citep{Planck2015:XIII}.}\label{tab:initial_condition_TT_planck}
	\begin{tabular}{c|c|c}
		\hline\hline
		Parameters & Minimum & Maximum  \\
		\hline
		$H_0$         & 57.7    & 76.9     \\
		$\Omega_bh^2$ & 0.0199  & 0.0245   \\
		$\Omega_ch^2$ & 0.0977  & 0.1417   \\
		$\tau$		  & 0.003   & 0.268    \\
		$10^9A_s$     & 1.4050  & 2.9861   \\
		$n_s$         & 0.9035  & 1.0275   \\
		\hline\hline
	\end{tabular}
\end{table}
\begin{figure*}
	\centering
	\includegraphics[width=0.9\textwidth]{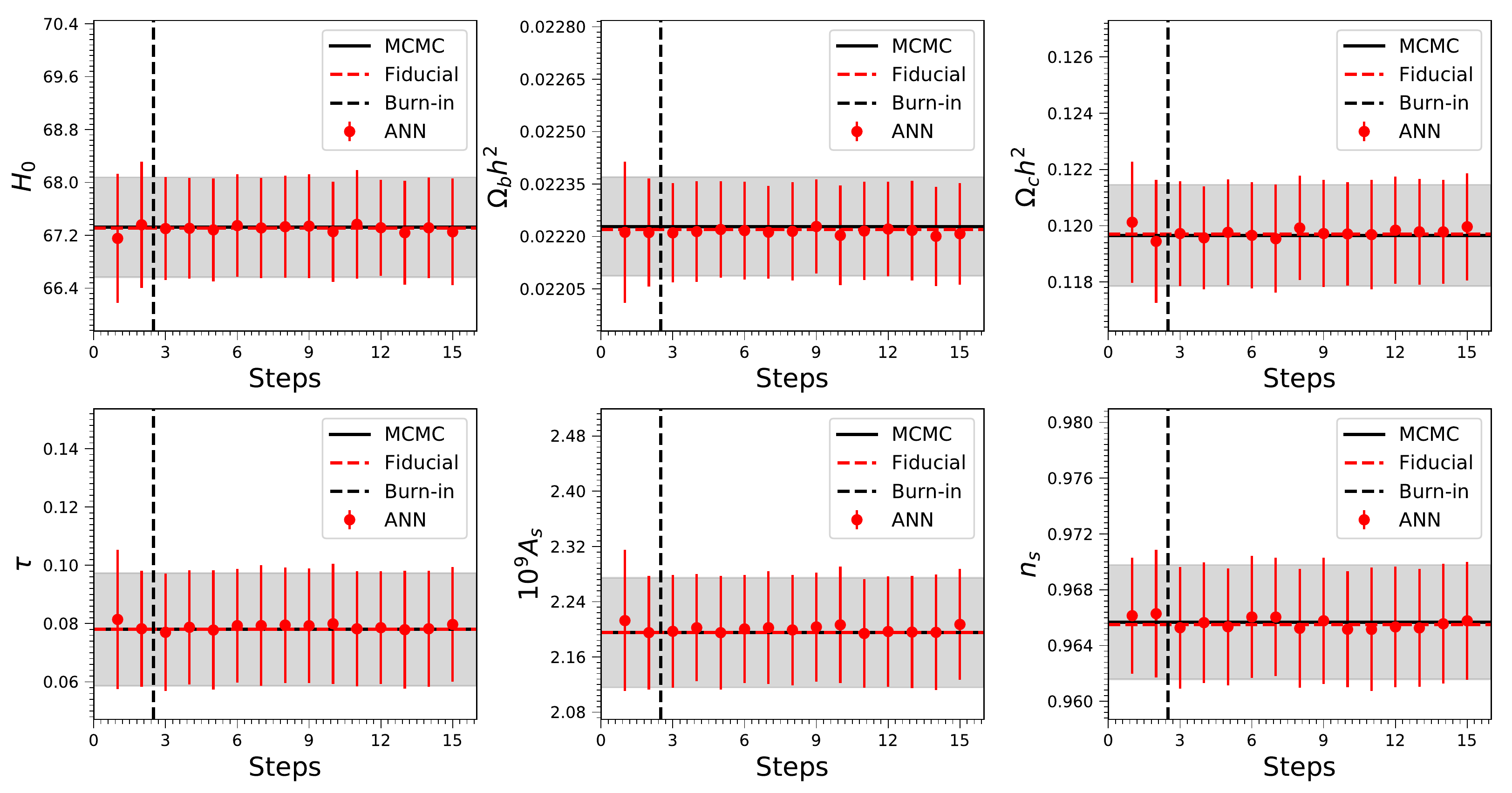}
	\caption{Same as Figure \ref{fig:steps_prism}, except now the initial conditions are based on the Planck2015 results.}\label{fig:steps_prism_2}
\end{figure*}

In the procedure of estimating parameters with an ANN, multiple chains of parameters will be obtained by training multiple ANNs (see section \ref{sec:training_process}). For the first ANN, it will be trained with samples simulated in the parameter space of the initial conditions. Then, a chain of parameters can be obtained by feeding the observational data to the ANN model, and the best-fit values and errors can be further calculated using this chain. Finally, these values of parameters will be used to update the parameter space to be learned by the ANN of the next step. In this way, after a limited number of steps, the parameter space will accurately cover the true values of parameters. The analysis of section \ref{sec:prism_predict_params} shows that the parameter space can be effectively updated at the end of each step, and the final parameter space can accurately cover the true values of parameters. This indicates that if the true values of parameters are not covered by the initial conditions, the ANN can cross the initial setting range of the parameters to find the true values of parameters. Therefore, the initial conditions are not factors of affecting the parameter estimation; thus, it can be set freely, which will be beneficial to models with insufficient prior knowledge of parameters. 

However, in order to reduce the training time, it is recommended to set large ranges of parameters for the initial conditions to ensure that the true values are covered. With the same procedure as section \ref{sec:prism_predict_params}, we estimate the six cosmological parameters using the temperature power spectrum of the PRISM CMB. Unlike the setting of $\sim 10\sigma$ biased initial conditions in section \ref{sec:prism_predict_params} (see Table \ref{tab:initial_condition_TT}), here we set good initial conditions that cover the fiducial values of cosmological parameters. Specifically, the initial conditions are set to $[P_{plk}-10\sigma_{plk}, P_{plk}+10\sigma_{plk}]$, where $P_{\rm plk}$ and $\sigma_{\rm plk}$ are the best-fit values and $1\sigma$ errors of the Planck2015 results \citep{Planck2015:XIII}. The setting of the initial conditions is shown in Table \ref{tab:initial_condition_TT_planck}, where the ranges of parameters are large enough to cover the fiducial cosmological parameters. Similarly, we obtain 15 chains by training 15 ANNs, and then we calculate the best-fit values and $1\sigma$ errors.

In Figure \ref{fig:steps_prism_2}, we show the best-fit values and $1\sigma$ errors of cosmological parameters as a function of steps. We can see that, at the first step, the parameters estimated by the ANN are consistent with the fiducial values within a $1\sigma$ confidence level, and the mean deviation between this result and the fiducial values is $0.144\sigma$. Furthermore, the results of the next 14 steps are stable and coincide with the fiducial values and the results of MCMC. Note that the burn-in phase contains only two steps, which is less than the steps in Figure \ref{fig:steps_prism}. Therefore, if good initial conditions are given to cover the true parameters, the ANN will be able to find the correct parameter space quickly, which can also reduce the time of parameter estimation with the ANN.

\subsection{Predict Multiple Experiments}

The strategy of adding noise illustrated in section \ref{sec:add_noise} allows multiple Gaussian noises to be added to samples of the training set, which ensure that the ANN learns not only the existing observation but also the observations with higher precision. Furthermore, the analyses of sections \ref{sec:test_higher_precision_mission} and \ref{sec:test_reliability_of_ANN_in_new_experiments} show that the ANNs trained with the current observation can also perform well for experiments that have 30\% uncertainties of the current experiments. Therefore, this means that the ANN can predict parameters not only for the current experiments but also for the future experiments that have higher precision, which means that multiple experiments can be learned by only one ANN. Therefore, this is beneficial for experiments that consume a lot of time and resources when estimating parameters.

Besides, the possibility of adding multiple noises to the training set means that the well-trained ANN can be used for parameter estimation in different stages of a specific experiment, and thus greatly reducing the time of parameter estimation. In addition, the method of adding multiple noises to the training set can also improve the robustness of the ANN in estimating parameters for the current observations, so that the parameters can be estimated with high accuracy.

\subsection{Time and Computing Resources}

With the increase of precision of experiments and the number of observational data, the consumption of time and computing sources in parameter estimation may be a problem to be solved for some experiments. Thus, it is very important to estimate cosmological parameters accurately and quickly. Fortunately, ECoPANN has this advantage in parameter estimation. 

Specifically, in the process of estimating parameters with an ANN, almost all the time is spent in the generation of the training sets and the training of the ANN, while very little time (about a few seconds) will be taken for estimating parameters with the trained ANN. In this work, ANNs are trained on one NVIDIA 1080 Ti graphics processing unit (GPU), and {\it emcee} is executed on two Intel Xeon E5-2690 v4 central processing units (CPUs) with a total of 28 cores. In the analysis of section \ref{sec:prism_predict_params}, 15 ANNs are trained totally to estimate parameters, which takes $\sim 126$ minutes. However, for the MCMC method, it takes $\sim 718$ minutes, which takes more time than the ANN method. For the joint constraint on parameters in section \ref{sec:joint_constraint}, eight ANNs are trained to estimate parameters, which takes $\sim840$ minutes, while for the MCMC method, it takes $\sim2418$ minutes.

\begin{figure}
	\centering
	\includegraphics[width=0.45\textwidth]{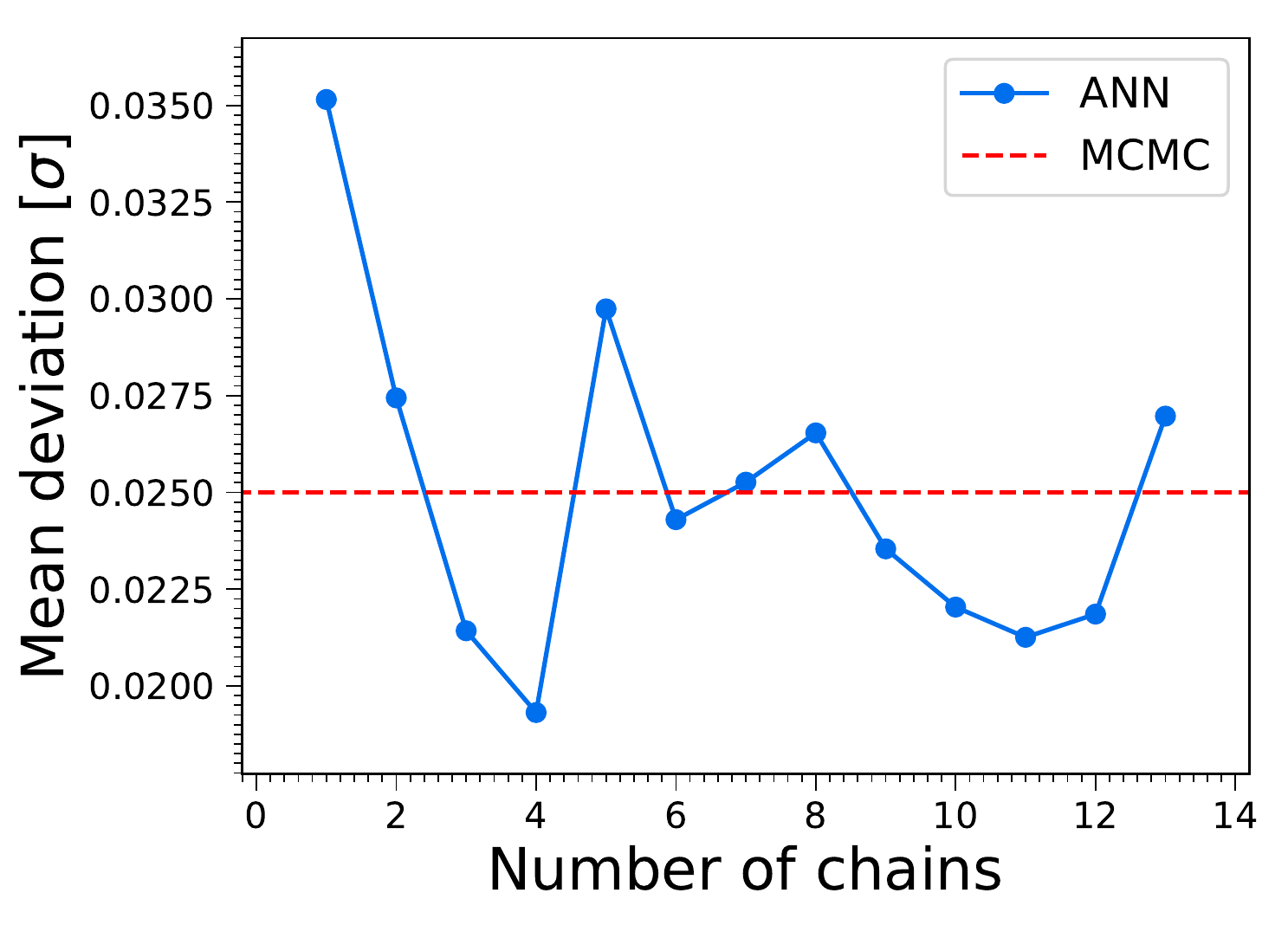}
	\caption{Mean deviations between the ANN results and the fiducial values as a function of the number of chains.}\label{fig:test_number_chains}
\end{figure}

Moreover, it should be noted that in the analysis of section \ref{sec:prism_predict_params}, $\sim 10\sigma$ biased initial conditions are given, which makes it spend a lot of time in the burn-in phase. However, the results of section \ref{sec:discussion_initial_conditions} (Figure \ref{fig:steps_prism_2}) show that if good initial conditions are given to cover the true values of cosmological parameters, the ANN will spend less time in the burn-in phase. In general, we can set large ranges for the initial conditions, and thus the burn-in phase will contain two steps, which means that the correct parameter space will be found after two steps. For the results of Figure \ref{fig:steps_prism_2}, there are 13 ANN chains can be used to estimate cosmological parameters. In order to test how many ANN chains are needed in the parameter estimation, we plot the mean deviations between the ANN results of Figure \ref{fig:steps_prism_2} and the fiducial values as a function of the number of chains, as shown in Figure \ref{fig:test_number_chains}. Despite that the maximum deviation when using one chain is $0.035\sigma$, it is similar to that of the MCMC method ($0.025\sigma$); thus, this should be acceptable in parameter estimation. Moreover, when using multiple chains, the deviations will be less than $0.030\sigma$, and all deviations are similar to those of the MCMC method. Therefore, for the ANN method, parameters can be estimated as long as there is one ANN chain. This means that, for many cases of parameter estimation that have two steps in the burn-in phase, we only need to train three ANNs, which will greatly reduce the time of estimating parameters.

Besides, from Figures \ref{fig:steps_prism} and \ref{fig:steps_prism_2}, we can see that the updated parameter space will intersect with the previous one, especially for steps after burn-in. Therefore, some samples in the training set can be reused in the next step to reduce time. Specifically, in the process of using ECoPANN, samples used in the previous step will be filtered according to the new parameter space, and then these selected samples will be used together with the newly generated samples to train the next ANN. Furthermore, the samples in the training set can also be saved to disk for further parameter estimations of the specific cosmological model. Therefore, when using ECoPANN, a sample database can be constructed for a specific cosmological model, which can reduce the time and computer resources spent on the repeated calculation of the model in parameter estimations. This will greatly facilitate the parameter estimation of time-consuming cosmological models.

In addition, the strategy of adding multiple noises (see section \ref{sec:add_noise}) makes it possible to use an ANN to estimate parameters for multiple experiments that have different precisions. Furthermore, we note that for the ANN used in section \ref{sec:prism_predict_params}, the time of generating 100,000 chains is $\sim4$ s, while for the multibranch network of section \ref{sec:joint_constraint}, it is $\sim19$ s. This means that in some cases one can estimate parameters in a few seconds with the well-trained ANNs directly, which will greatly reduce the time of parameter estimation. Therefore, these advantages of ANNs can greatly reduce the consumption of time and computing resources of parameter inference, which may be very beneficial to the current and future experiments.

\subsection{Covariance Matrix}

It should be noted that covariance between the measurements is not considered in the analysis of sections \ref{sec:application_CMB} and \ref{sec:joint_constraint}. However, this does not mean that our method is not capable of dealing with observational data sets that have a covariance matrix. To do this, the key is to change the type of noise added to the training set. In the strategy of section \ref{sec:add_noise}, random Gaussian noise is added to the training set without considering the correlation between the measurements. Therefore, in order to consider the correlation between the measurements, noise that is subjected to multivariate Gaussian distribution should be added to the training set. Specifically, in the training process, noise subjected to $\mathcal{N}(0,\Sigma)$ is generated and added to each sample of the training set, where $\Sigma$ is the covariance matrix of the observational data. Note that the noise added to the training set depends on the specific experimental observation. This means that for observational data with covariance, the well-trained ANN may not be able to estimate parameters for higher-precision experiments. We will study this issue further in our future work.

To test the capability of the ANN in dealing with observational data sets that have a covariance matrix, we constrain $w$ and $\Omega_m$ of the $w$CDM model using the latest Pantheon SNe Ia \citep{Scolnic:2018}. The Pantheon SN Ia data contain 1048 data points within the redshift range of [0.01, 2.26]. The distance modulus of Pantheon SNe Ia can be rewritten as
\begin{equation}
\mu=m_{B,corr}^* - M_{B} ~,
\end{equation}
where $m_{B,corr}^* = m_{B}^*+\alpha\times x_1-\beta\times c + \Delta_B$ are the corrected apparent magnitudes reported in \citet{Scolnic:2018}, and $M_B$ is the absolute magnitude of the $B$ band. Since the absolute magnitude $M_B$ of SNe Ia is strongly degenerate with the Hubble constant $H_0$, we combine $M_B$ and $H_0$ to be a new parameter and constrain it with the cosmological parameters simultaneously. In our analysis, the systematic uncertainties are considered in estimating parameters, and thus the systematic covariance matrix $\bf C_{\rm sys}$ is used in the process of adding noise to the training set. We note that the measurement of Pantheon SNe Ia is the corrected apparent magnitudes; thus, the input of the ANN is $m_{B,corr}^*$ (or $\mu + M_B$ generated by the $w$CDM model).

\begin{figure}
	\centering
	\includegraphics[width=0.45\textwidth]{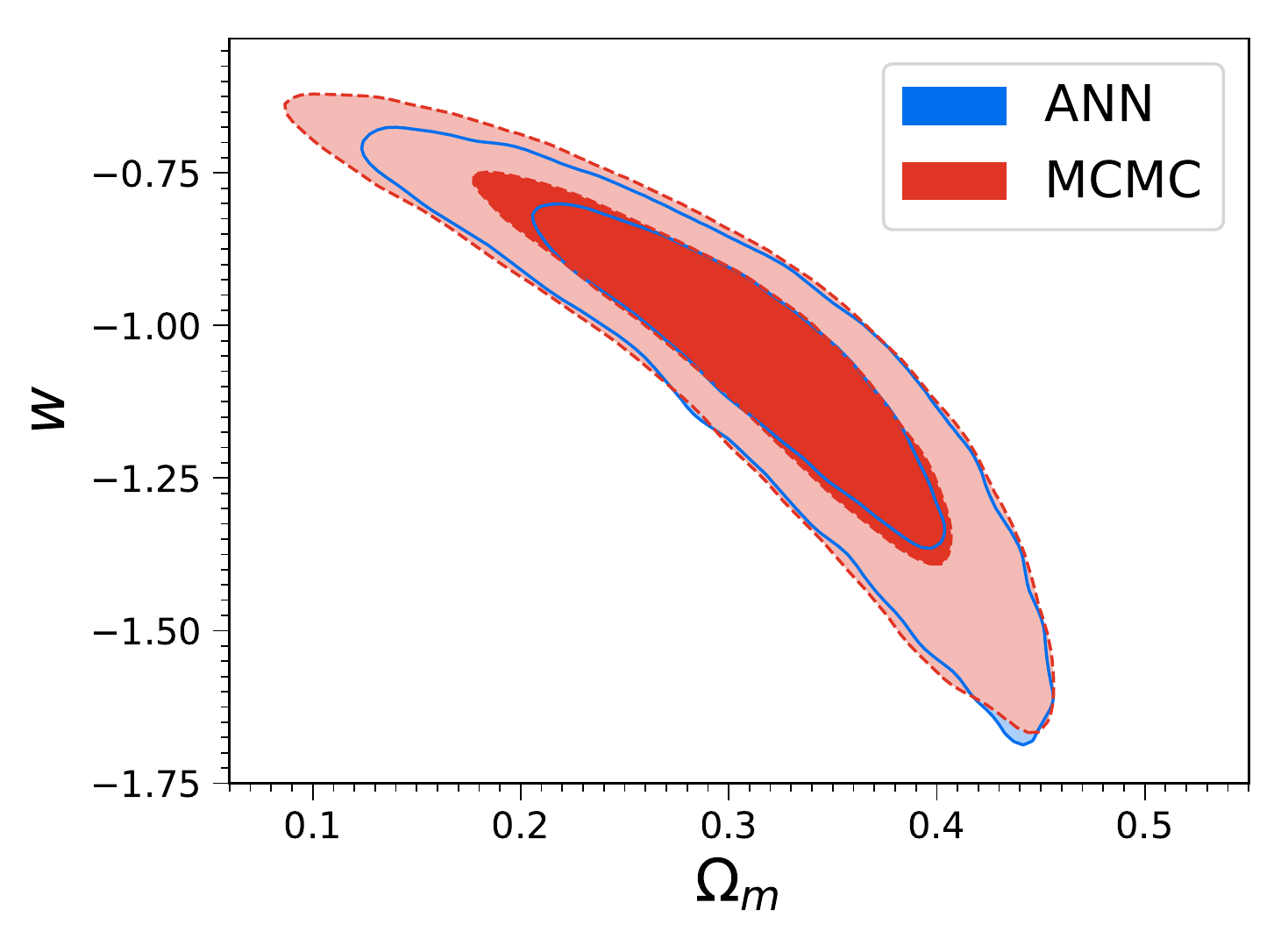}
	\caption{Two-dimensional marginalized distributions with $1\sigma$ and $2\sigma$ contours of $w$ and $\Omega_m$ of the $w$CDM model constrained from Pantheon SNe Ia with systematic uncertainties.}\label{fig:wCDM_Cov}
\end{figure}

In our analysis, 10 ANNs are trained to estimate cosmological parameters, and the time consumed is $\sim192$ minutes, which is also less than that of the MCMC method ($\sim594$ minutes). In Figure \ref{fig:wCDM_Cov}, we show the two-dimensional distributions of $w$ and $\Omega_m$. For the ANN method, the best-fit values with $1\sigma$ errors are
\begin{align}
w&=-1.052_{-0.213}^{+0.162}, & \Omega_m&=0.318_{-0.067}^{+0.062},
\end{align}
and for the MCMC method, the best-fit values with $1\sigma$ errors are
\begin{align}
w&=-1.047_{-0.228}^{+0.193}, & \Omega_m&=0.317_{-0.083}^{+0.064}.
\end{align}
We can see that the results of the ANN method and the MCMC method are consistent with the results of \citet{Scolnic:2018} within a $1\sigma$ confidence level. Furthermore, for the two parameters, the deviations between the ANN results and the MCMC results are $0.018\sigma$ and $0.013\sigma$, which are small enough to be acceptable in parameter estimations. The results show that the ANN method can correctly obtain the best-fit values, errors, and correlations of parameters. Therefore, the ANN method is capable of dealing with observational data sets that have covariance matrices.

\begin{figure}
	\centering
	\includegraphics[width=0.45\textwidth]{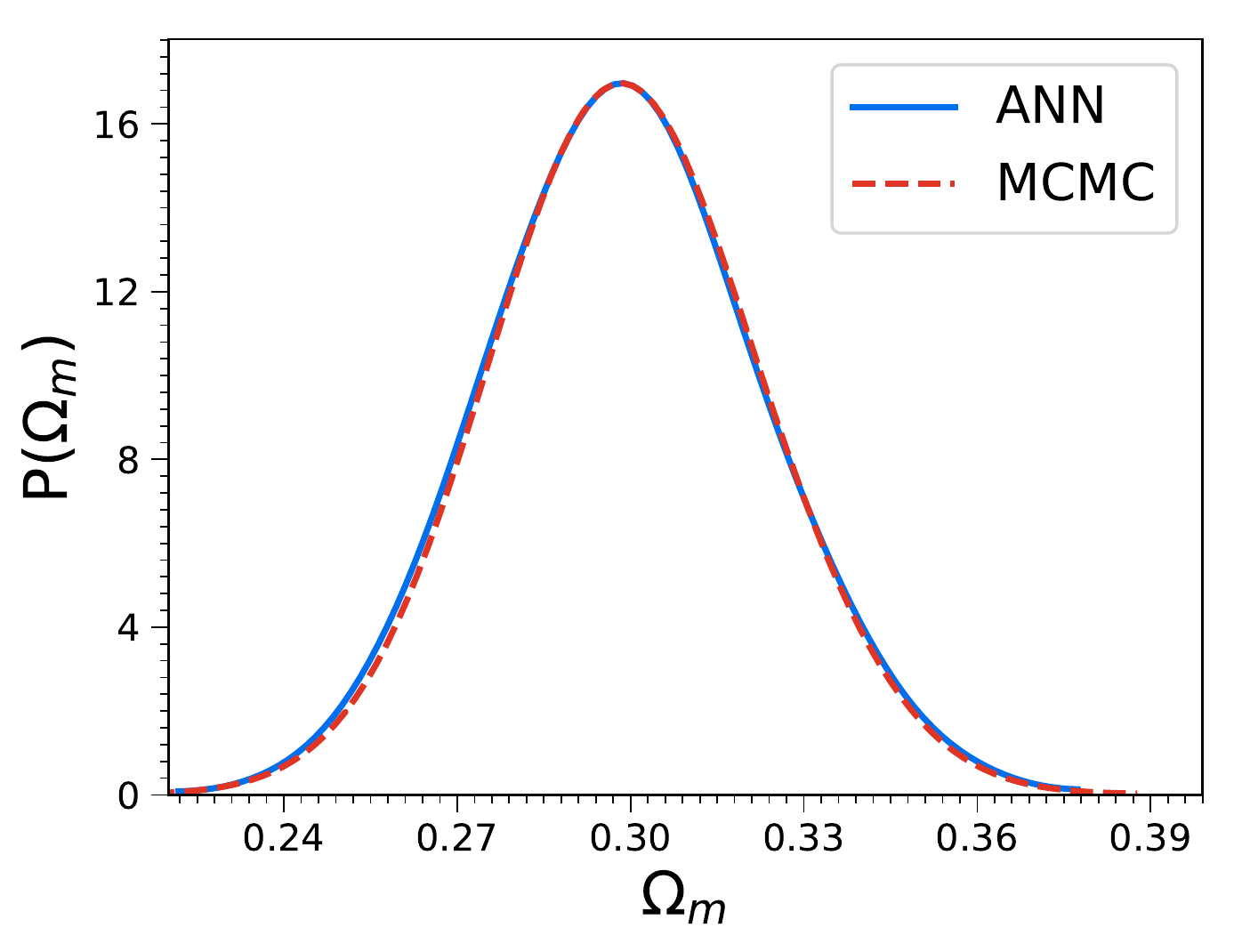}
	\caption{One-dimensional marginalized distributions of $\Omega_m$ constrained from Pantheon SNe Ia with systematic uncertainties.}\label{fig:LCDM_Cov}
\end{figure}

The contours in Figure \ref{fig:wCDM_Cov} show that the ANN gives slightly tighter constraints than MCMC. After careful check, we conclude that this may be caused by the parameter space of $\Omega_m$ learned by the ANN. Specifically, as shown in Figures \ref{fig:test_biased_H0} and \ref{fig:test_biased_all_params}, when the true parameters deviate from the median of the parameter space by more than $3\sigma_p$, the estimated parameters will deviate slightly from the true values. In the training process of the ANN, the parameter space of $\Omega_m$ will be cut off by 0, which makes it impossible to learn a larger parameter space. Therefore, for measurements where $\Omega_m$ is close to 0, poor parameter values may be obtained. To further test this, we set $w=-1$ and constrain $\Omega_m$. The one-dimensional marginalized distributions of $\Omega_m$ are shown in Figure \ref{fig:LCDM_Cov}. For the ANN method, the best-fit value and $1\sigma$ error are
\begin{equation}
\Omega_m = 0.299\pm0.023,
\end{equation}
and for the MCMC method, the best-fit value and $1\sigma$ error are
\begin{equation}
\Omega_m = 0.299\pm0.022.
\end{equation}
Obviously, the results of the ANN method are almost the same as those of the MCMC method. In this case, the parameter space learned by the ANN is not cut off by $\Omega_m=0$; therefore, it is possible for the ANN to learn a large enough parameter space to accurately estimate the parameters.

It should be noted that for future SN Ia data (with more data points or higher precision), the result of Figure \ref{fig:wCDM_Cov} will be better. To solve a problem like Figure \ref{fig:wCDM_Cov}, we will further study how to better learn the parameter space in our future work.

\section{\bf Conclusions}\label{sec:conclusions}

In this work, we present a new method to estimate cosmological parameters accurately using an ANN. Based on ANN, a framework called ECoPANN is developed to achieve parameter inference, which can be used on CPUs or GPUs. Our analysis shows that the well-trained ANN model performs excellently on both the best-fit values and errors, as well as correlations between parameters when compared with that of the traditional MCMC method. More importantly, ECoPANN has advantages in parameter estimation. Specifically, the initial conditions of parameters can be set more freely. This means that the true parameter will be obtained even when biased initial conditions are given, which is beneficial to models with insufficient prior knowledge of parameters. Furthermore, the strategy of adding noise in ECoPANN makes it possible to use an ANN to predict parameters for multiple experiments that have different precisions. Moreover, the ECoPANN is designed to reduce the consumption of time and computing resources by reusing samples of the cosmological model. Therefore, when using ECoPANN, a sample database can be constructed for a specific cosmological model, which will greatly facilitate the parameter estimation of time-consuming cosmological models. These advantages of ANNs may give them more potential than the MCMC method in parameter inference.

In addition to estimating parameters with one observational dataset, we also expand the ANN model to a multibranch network to achieve a joint constraint on parameters using multiple observational data sets. We test the multibranch network with the simulated CMB, SN Ia, and BAO data sets, and the results show that the multibranch network also performs well in parameter estimation. Therefore, the ANN method is capable of estimating parameters using data sets of multiple experiments in the future.

ANNs provide an accurate and fast alternative to the MCMC method that is commonly used by researchers in astronomy. Their effectiveness in analyzing one-dimensional curve data proves them to be a general method that can be used for parameter estimation in many experiments to facilitate research in cosmology and even other, broader scientific fields.

\section{\bf Acknowledgement}

We thank Jun-Feng Li, Xu Li, Jie Liu, Xiao-Jiao Ma, Yang Yang, and Ji-Ping Dai for helpful discussions. We thank Heng Yu for useful suggestions. J.-Q.X. is supported by the National Science Foundation of China under grant Nos. U1931202, 11633001, and 11690023 and the National Key R\&D Program of China No. 2017YFA0402600.

\end{document}